\documentclass[onecolumn]{IEEEtran}
\usepackage[T1]{fontenc}
\usepackage[latin9]{inputenc}
\usepackage{float}
\usepackage{amsmath}
\usepackage{amsthm}
\usepackage{amssymb}
\usepackage{graphicx}
\usepackage{setspace}
\usepackage[unicode=true,
 bookmarks=true,bookmarksnumbered=true,bookmarksopen=true,bookmarksopenlevel=1,
 breaklinks=false,pdfborder={0 0 0},pdfborderstyle={},backref=false,colorlinks=false]
 {hyperref}
\hypersetup{
 hypertexnames=false}

\makeatletter

\floatstyle{ruled}
\newfloat{algorithm}{tbp}{loa}
\providecommand{\algorithmname}{Algorithm}
\floatname{algorithm}{\protect\algorithmname}

\theoremstyle{plain}
\newtheorem{thm}{\protect\theoremname}
\theoremstyle{definition}
\newtheorem{defn}[thm]{\protect\definitionname}
\theoremstyle{plain}
\newtheorem{lem}[thm]{\protect\lemmaname}

\usepackage{cite}

\makeatother

\providecommand{\definitionname}{Definition}
\providecommand{\lemmaname}{Lemma}
\providecommand{\theoremname}{Theorem}

\begin{document}

\title{Distributed Learning for Channel Allocation Over a Shared Spectrum}

\author{S.M. Zafaruddin,~\IEEEmembership{Member,~IEEE}, Ilai~Bistritz,~\IEEEmembership{Student Member,~IEEE},
Amir~Leshem,~\IEEEmembership{Senior Member,~IEEE} and Dusit (Tao)
Niyato~\IEEEmembership{Fellow,~IEEE}\thanks{S.\textasciitilde M.\textasciitilde{} Zafaruddin was with the Faculty
of Engineering, Bar-Ilan University, Ramat Gan 5290002, Israel (e-mail:
smzafar@biu.ac.il). Currently, he is with the Department of Electrical
and Electronics Engineering, BITS Pilani, Pilani-333031 (email: syed.zafaruddin@pilani.bits-pilani.ac.in).
I. Bistritz is with the Faculty of Engineering, Bar-Ilan University,
Ramat Gan 5290002, Israel (e-mail: ilaibist@gmail.com). A. Leshem
is with the Faculty of Engineering, Bar-Ilan University, Ramat Gan
5290002, Israel (e-mail: leshema@eng.biu.ac.il). Dusit (Tao) Niyato
is with School of Computer Science and Engineering (SCSE), Nanyang
Technological University, Singapore 639798. (e-mail: dniyato@ntu.edu.sg).\protect \\
This research was supported by the ISF-NRF Joint research Program,
under grant ISF 2277/16. S. M. Zafaruddin was partially funded by
the Israeli Planning and Budget Committee (PBC) post-doctoral fellowship. }}
\maketitle
\begin{abstract}
Channel allocation is the task of assigning channels to users such
that some objective (e.g., sum-rate) is maximized. In centralized
networks such as cellular networks, this task is carried by the base
station which gathers the channel state information (CSI) from the
users and computes the optimal solution. In distributed networks such
as ad-hoc and device-to-device (D2D) networks, no base station exists
and conveying global CSI between users is costly or simply impractical.
When the CSI is time varying and unknown to the users, the users face
the challenge of both learning the channel statistics online and converge
to a good channel allocation. This introduces a multi-armed bandit
(MAB) scenario with multiple decision makers. If two users or more
choose the same channel, a collision occurs and they all receive zero
reward. We propose a distributed channel allocation algorithm that
each user runs and converges to the optimal allocation while achieving
an order optimal regret of $O\left(\log T\right)$. The algorithm
is based on a carrier sensing multiple access (CSMA) implementation
of the distributed auction algorithm. It does not require any exchange
of information between users. Users need only to observe a single
channel at a time and sense if there is a transmission on that channel,
without decoding the transmissions or identifying the transmitting
users. We demonstrate the performance of our algorithm using simulated
LTE and 5G channels.
\end{abstract}


\section{Introduction}

Channel allocation in wireless communication is one of the fundamental
management tasks. It and has been widely studied for various wireless
networks \cite{Katzela2000,Chieochan2010,Ku2015,Tragos2013,Tanab2017}.
In the traditional centralized systems, Orthogonal Frequency Division
Multiplexing Access (OFDMA) was investigated extensively to meet the
high demand for efficient spectrum utilization. If users can be assigned
to sub-channels efficiently, certain gains can be derived from the
diversity of the channel. The main issue for the OFDMA systems is
joint power and sub-carrier allocation in the downlink direction \cite{Wong1999,Shen2005,Sadr2009,Huberman2012}
and sub-carrier assignment in the uplink direction \cite{Gao2008,Huang2009,Yaacoub2012}.
Due to the global view of the whole network, the centralized approach
is able to obtain the optimal solution of a desired performance metric.
The optimal channel allocation can be computed using the well-known
Hungarian method \cite{Papadimitriou1998}.

However, there are some disadvantages that limit the practicality
of the centralized approach such as significant signaling overhead,
increased implementation complexity and higher latency in dealing
with resource allocation problems. Moreover, emerging wireless networking
paradigms such as cognitive radio networks, ad-hoc networks, and D2D
communications are inherently distributed. A complete information
about the network state is typically not available online, which makes
the computation of optimal policies intractable for these networks.
Hence, it is desirable to develop a distributed learning algorithm
for dynamic spectrum access that can effectively adapt for general
complex real-world settings in dense and heterogeneous wireless environments.

Open sharing model employs spectrum sharing among peer users as the
basis for managing a spectral band. Advocates of this model draw support
from the phenomenal success of wireless services operating in the
unlicensed industrial, scientific, and medical (ISM) radio band (e.g.,
WiFi). Centralized and distributed spectrum sharing strategies have
been initially investigated to address technological challenges under
this spectrum management model.

The center of the channel allocation task is the combinatorial optimization
assignment problem. Solving the assignment problem distributively
is a major challenge that has received considerable attention. The
famous auction algorithm \cite{bertsekas1979distributed} proposed
a distributed method to solve the assignment problem where users send
their bids to an auctioneer. In \cite{naparstek2013optimal} a fully
distributed version of the auction algorithm was suggested that exploits
carrier sense multiple access (CSMA) in order to avoid the need for
an auctioneer.

If the resources (channels) values are not known in advance by the
users, they have to learn these values online. Learning the CSI in
real-time comes at the expense of using the best known channels so
far. This introduces the well-known trade off between exploration
and exploitation, that is captured by the multi-armed bandit (MAB)
problem. In this case, there are several decision makers facing this
problem, and when two or more choose the same channel, they receive
zero reward. Similarly to other MAB problems, the performance is measured
by the expected difference between the actual sum of rewards and the
sum of rewards that could have been achieved if the users had perfect
knowledge of the CSI. However, as opposed to classical MAB problems,
the interaction between the users significantly complicates the learning
aspects of the problem. To address that, deep reinforcement learning
and Q-learning methods have been proposed for these problems \cite{Challita2017,Wang2018,Naparstek},
and have been shown to perform well for small-size models. However,
for large-scale networks these methods perform poorly since the number
of states of the learning algorithm increases exponentially in the
number of users.

In \cite{Nayyar2016}, the auction algorithm \cite{bertsekas1979distributed}
was used as a basis for a distributed algorithm that achieves an expected
sum regret of $O(\log T)$. However, since it relies on \cite{bertsekas1979distributed}
, this algorithm requires communication between users in order to
communicate the bids and deduce which player won each auction. To
implement this algorithm, users need to know which user transmitted
on which channel. In this manner, they can use their public channel
choices as a signaling method. In practice, this knowledge requires
that users decode at least part of the transmission to identify the
ID of the transmitting users. Besides being computationally demanding,
this might be highly non-trivial when multiple users transmit on the
same channel and all their IDs need to be decoded from the mixture.

In this paper, we overcome this requirement by proposing a distributed
algorithm that relies on \cite{naparstek2013optimal} instead of \cite{bertsekas1979distributed}.
The algorithm in \cite{naparstek2013optimal} assumes that the CSI
is known. It also uses a continuous back-off time and assumes no tied
bids. We lift all of these assumptions in our novel MAC protocol.
Our protocol achieves an expected sum of regret of $O(\log T)$, but
in contrast to \cite{Nayyar2016}, only requires each user to sense
the channel that the user is using and detect if there are other transmissions
on this channel. Users do not need to know which user transmitted
on which channel or how many of them did. Therefore, our algorithm
offers the same order optimal performance as \cite{Nayyar2016} but
with dramatically simpler implementation.

\subsection{Related Works}

Developing multi-armed bandit (MAB)-based methods for solving dynamic
spectrum allocation (DSA) problems is a relatively new research direction,
motivated by recent developments of MAB in various other fields, and
many works have been done in this direction recently. A couple of
these works \cite{Avner2014,Anandkumar2011,Liu2012,Liu2010Restless}
considered a cognitive radio scenario where a set of channels can
be either empty or occupied by a primary user that interferes all
secondary users. A generalized scenario was considered in \cite{Vakili2013,Rosenski2016,Liu2010MAB},
where the channel qualities are not binary, but still all users have
the same vector of channel qualities. Recently, the case of a full
channel allocation scenario where different users have different channel
qualities (a matrix of channel qualities) have been considered in
\cite{Kalathil2014}, and later improved in \cite{Nayyar2016}, by
the same authors, to have an order optimal sum-regret of $O\left(\log T\right)$.

Recently, it has been shown in \cite{Bistritz} (which improved \cite{bistritz2018distributed})
that achieving a sum-regret of near-$O\left(\log T\right)$ is possible
even without communication between users and with a matrix of expected
rewards. The algorithm in \cite{Bistritz} is general and has a slow
convergence rate in $T$ that makes it unsuited for realistic communication
scenarios. In this paper, we adopt a more practical and communication
oriented approach and achieve an order optimal sum-regret of $O\left(\log T\right)$.
Our algorithm still does not require any communication between users,
and each device only needs to sense a single channel at a time (instead
of simultaneously all of them as in \cite{Nayyar2016}). It is made
possible by adding assumptions that are always valid from a practical
perspective - the expected rewards (QoS) are integer multiplications
of a common resolution $\Delta_{\min}$, and a device can choose not
to transmit on any channel and instead only to sense a single channel
of its choice. Our algorithm is much easier and less costly to implement
than that of \cite{Nayyar2016} and has a much better convergence
time that that of \cite{Bistritz}.

The literature on distributed channel allocation without learning,
where the CSI is assumed to be known, is vast and we can only cover
part of it here. Recently there has been growing interest in distributed
spectrum optimization for frequency selective channels, where the
assignment problem arises. However, most of the work done in this
field relies on explicit exchange of CSI. Several suboptimal approaches
that do not require information sharing have been suggested. In \cite{Kwon2010},
a greedy approach to the channel assignment problem was introduced.
In \cite{Yaffe2010} and \cite{Leshem2012}, the use of opportunistic
carrier sensing was combined with the Gale-Shapley algorithm for stable
matching \cite{gale1962college} to provide a fully distributed stable
channel assignment. This solution basically achieves the greedy channel
assignment and analysis of this technique for Rayleigh fading channels
was done in \cite{Naparstek2012}.

Game theory is often used to design distributed channel allocation
algorithms. In \cite{Mochaourab2015} the channel assignment problem
was formulated as a many-to-one matching game under the limitation
that each primary channel can only be assigned to one secondary user.
In \cite{Bistritz2018}, an algorithm was proposed based on a game
with utility design that leads to an asymptotically optimal performance
in all Nash equilibria. In \cite{Xiao2015} the spectrum sharing problem
between D2D pairs and multiple co-located cellular networks was formulated
as a Bayesian non-transferable utility overlapping coalition formation
game. Nash bargaining solutions for channel allocation were considered
in \cite{leshem20112,Han2005,Leshem2011}, and distributed allocation
using multichannel ALOHA and potential games was considered in \cite{Cohen2013Journal,Cohen2013}.

The auction algorithm has been extensively used to solve a variety
of assignment problems. It gets its name from operating similarly
to an auction. As in this paper and many others, the auction algorithm
may have nothing to do with actual auctions that rely on economic
and game-theoretic principles, as was done in \cite{Sun2006,Han2011,Mukherjee2010,Chang2010}.
In \cite{Yang2009} the auction algorithm was used to solve the channel
assignment problem for the uplink, using the base station as the auctioneer.
In \cite{Bayati2007} a distributed auction algorithm with shared
memory was used for switch scheduling. In \cite{Bayati2008} it was
shown that a modification of the auction algorithm is equivalent to
max product belief propagation. However, all these modified auction
algorithms require a base station or shared memory, which prevents
them from being fully distributed. In addition, all these algorithms,
including \cite{naparstek2013optimal} that is being used here, assume
that the CSI is known to the users. Our algorithm generalizes the
distributed CSMA auction algorithm \cite{naparstek2013optimal} to
an online learning framework.

\subsection{Outline}

The paper is organized as follows: in Section II we describe the system
model and our network assumptions. Section III discusses the novel
MAC protocol we propose. Section IV and Section V analyze the exploration
phase and auction phases of our algorithm, respectively. Section VI
provides simulation results of our algorithm on practical LTE channels,
together with a performance comparison. Finally, Section VII concludes
the paper.

\section{System Model\label{section:system}}

We consider an Ad-Hoc network with a set of transmitter-receiver pairs
(links) $\mathcal{N}=\left\{ 1,\ldots,N\right\} $ and a set of channels
$\mathcal{K}=\left\{ 1,\ldots,K\right\} $, where $K\geq N$. Each
channel consists of several OFDMA subcarriers and each link uses a
single channel. In the case of more users than channels ($N>K$),
a combined OFDMA-TDMA can be used instead in order to have enough
resources for all users. However, since this is a trivial consequence
of our analysis which only complicates the notation, we choose to
avoid considering TDMA. The number of channels $K$ is chosen by the
protocol designer to be large enough to support $N$ links in an environment
with outside interferers where some of the channels can be very poor
and practically unavailable. The identity and number of subcarriers
that constitute each channel can also be optimized with respect to
the typical channels used by the significant interferes. Links may
use multiple-input multiple-output (MIMO) transmission, with different
capabilities for each link. Time is slotted and indexed by $t$, such
that in each time slot $L$ OFDM symbols are transmitted. The number
of OFDM symbols per time slot $L$ can be designed to match the coherence
time of the channel, such that the CSI typically changes every time
slot. Hence, we assume a fast-fading scenario where the coherence
time is proportional to an OFDM symbol duration. The links are active
for a total of $T$ time slots, where $T$ is unknown in advance by
the links. We assume that each link can sense a single channel at
each time slot, which is the channel they use, and detect whether
other links are transmitting on this channel. The chosen channel of
link $n$ at time $t$ is denoted by $a_{n}\left(t\right)$. Naturally,
links can choose not to transmit at all at a given time slot, which
is denoted $a_{n}\left(t\right)=0$. Non-transmitting links can still
sense transmissions on a single chosen channel.

\begin{figure}[t]
~~~~~~~~~~~~~~~~~~~~~~~~~~~~~~~~~~~~~~~~~~\includegraphics[width=7.5cm,height=7.5cm]{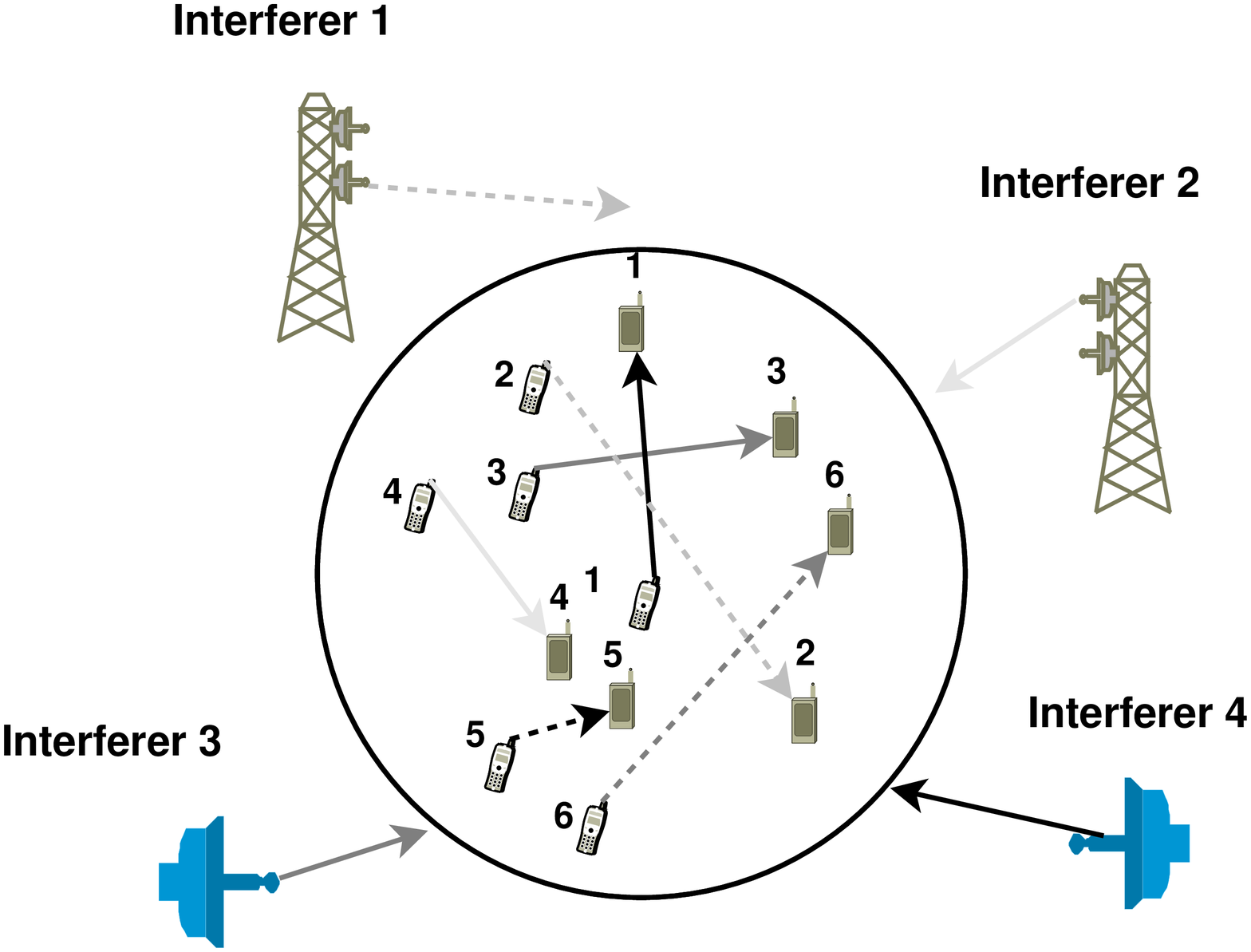}

\caption{\label{fig:System-Model}System Model}
\end{figure}

The links are located in a geographical proximity in an area that
typically includes other coexisting networks nearby. As a result,
each receiver experiences alien interference from the transmission
of other protocols. Due to the geometry of the links and the different
channels used by different interferers, the average interference is
different for each receiver in our network. A toy example of our network
with $K=N=6$ is depicted in Fig. \ref{fig:System-Model}. The channel
used by each link is indicated by the color of the arrow between its
transmitter and receiver. Outside the area of the network there are
four major interferers that use four of the six available channels.
In this example, links successfully avoid using channels with significant
interference at their receiver side.

This outside transmissions can be constant over time or bursty, and
may overlap any part of the subcarriers used by a particular link.
In addition, the fading of the channel may cause significant changes
to the channel gains of the subcarriers. As any modern device, the
transmitter and receiver of each link adopt techniques such as adaptive
beamforming and modulation together with interleaving and coding for
fading channels in order to provide a stable (on average) and reliable
communication for the users. However, since the channel statistics
and the interference pattern are initially unknown, each link needs
to learn them online as fast as possible in order to deduce which
Quality of Service (QoS) it can support. As in any practical system,
there is some resolution for the supported QoS (e.g., 100Kbps), we
denote by $\Delta_{\min}$. The supported QoS set is $\mathcal{Q}\triangleq\left\{ Q_{1},\ldots,Q_{M}\right\} $
where for each $i$, $Q_{i}=l_{i}\Delta_{\min}$ for a non-negative
integer $l_{i}$ and $Q_{1}<\ldots<Q_{M}$. The QoS experienced by
link $n$ using channel $i$ is denoted by $Q_{n,i}$. A value in
this set may represent the weighted quality of a combination of parameters,
e.g., 1Mbps for internet, 256kbps for voice and 10Mbaps for video.
In general, different links have a subset of different possible QoS
values from $\mathcal{Q}$ due to different capabilities, e.g., number
of transmitting and receiving antennas. Being part of the standard
of the protocol, we assume that the parameters $\Delta_{\min}$ and
$\Delta_{\max}=Q_{M}-Q_{1}$ are known to all links.

In each time slot $t$, each link measures the instantaneous QoS $q_{n,i}\left(t\right)$
by using a finer resolution than that of $\mathcal{Q}$, in order
for the estimation of the average to be accurate. We model $q_{n,i}\left(t\right)$
as an i.i.d. sequence in time, independent for different $n$ or $i$.
The distribution of $q_{n,i}\left(t\right)$ is bounded since $Q_{1}\leq q_{n,i}\left(t\right)\leq Q_{M}$,
and can be either discrete or continuous due to arbitrarily fine measurements.

Define the set of links that are transmitting on channel $i$ at time
$t$ by
\begin{equation}
\mathcal{N}_{i}\left(t\right)=\left\{ n\,|\,a_{n}\left(t\right)=i\right\} .\label{eq:1}
\end{equation}
Define the no-collision indicator of channel $i$ at time $t$ by
\begin{equation}
\eta_{i}\left(t\right)=\Biggl\{\begin{array}{cc}
0 & \Bigl|\mathcal{N}_{i}\left(t\right)\Bigr|>1\\
1 & o.w.
\end{array}.\label{eq:2}
\end{equation}
The instantaneous reward of link $n$ at time $t$ from transmitting
on channel $a_{n}$ is

\begin{equation}
r_{n,a_{n}}\left(t\right)=q_{n,a_{n}}\left(t\right)\eta_{a_{n}}\left(t\right).\label{eq:3}
\end{equation}
The theoretical guarantees of our algorithm are formulated using the
well-known notion of regret, defined as follows.
\begin{defn}
The total regret is defined as the random variable
\begin{equation}
R=\sum_{t=1}^{T}\sum_{n=1}^{N}Q_{n}^{*}-\sum_{t=1}^{T}\sum_{n=1}^{N}q_{n,a_{n}\left(t\right)}\left(t\right)\eta_{a_{n}\left(t\right)}\left(t\right).\label{eq:4}
\end{equation}
 The value $Q_{n}^{*}$ is the expectation of the QoS of the channel
link $n$ is assigned to in
\begin{equation}
a^{*}=\arg\underset{a_{1},...,a_{N}}{\max}\sum_{n=1}^{N}Q_{n,a_{n}}.\label{eq:5}
\end{equation}
The expected total regret $\bar{R}\triangleq E\left\{ R\right\} $
is the average of \eqref{eq:4} over the randomness of the rewards
$\left\{ r_{n,i}\left(t\right)\right\} _{t}$, that dictate the random
channel choices $\left\{ a_{n}\left(t\right)\right\} $.
\end{defn}

\section{Protocol Description\label{section:protocol}}

We design a novel MAC protocol that each link runs distributedly in
order to maximize the accumulated sum of QoS. In the original auction
algorithm, an auctioneer is needed to collect the bids and compute
the highest bidder. Such an auctioneer is not available in a distributed
wireless network. The algorithm in \cite{naparstek2013optimal} exploits
the CSMA mechanism to bypass the need for an auctioneer and by doing
that, implements the auction algorithm distributedly. For this purpose,
links compute a continuous back-off time that is decreasing with their
bid. The highest bidder for a particular channel is simply the first
link that accesses this channel. Since we assume all links can sense
the channel they chose, all links will agree on which link was the
highest bidder for their channel. \textbf{Note that we are not analyzing
selfish links but devices that are programmed to run our designed
MAC protocol. }

The key advantage of our algorithm is that it only requires from each
receiver to sense if there are transmissions on a single channel,
which is a basic requirement. We assume that all links are of a sensing
distance from each other (a fully-connected network). However, as
opposed to \cite{Nayyar2016}, links do not know which transmission
belongs to which link. This is the scenario in practice with wireless
links located in close enough proximity. In our protocol, links do
not need to distinguish between the transmission of other links, which
might have required decoding an ID for each link. Moreover, it can
be extremely demanding in practice to separate colliding transmissions
and discern the IDs involved. Sensing a single channel at a time instead
of all the $K$ channels is another major advantage of our algorithm
over \cite{Nayyar2016}.
\begin{defn}
We divide the $T$ time slots into packets with a dynamic length,
one starting immediately after the other. Each packet is further divided
into three phases. In the $k$-th packet:
\end{defn}
\begin{enumerate}
\item \textbf{Exploration Phase} - this phase has a length of $c_{1}$ time
slots in each packet, and is used for estimating the expected reward
in each channel. The estimated values are artificially dithered in
order to avoid ties in the subsequent auction phase. This phase is
described in detail and analyzed in Section \ref{sec:Pure-Exploration-Phase}.
It adds a $O\left(\log T\right)$ to the expected total regret.
\item \textbf{Auction Phase} - this phase has a length of $\left\lceil 4K^{2}N\left(\frac{Q_{M}}{\Delta_{\min}}+\frac{1}{N}\right)\left(2^{b\left(k\right)}+1\right)\right\rceil $
time slots in the $k$-th packet, which is the convergence time of
the distributed auction algorithm, as dictated by Lemma \ref{lem:AuctionConvergence}.
In this phase, links run the distributed auction on the estimated
expected rewards using $b\left(k\right)$ bits for the quantized back-off
time. The function $b\left(k\right)$ converges to a constant independent
of $k$. In practice, it is easy to guarantee that $b\left(0\right)$
is already large enough, but the designer can shorten the convergence
time by starting from smaller $b\left(0\right)$ values and let the
algorithm find the minimal $b\left(k\right)$ necessary. This phase
is analyzed in detail in Section \ref{sec:Auction Phase}.
\item \textbf{Exploitation Phase} - this phase has a length of $c_{2}2^{k}$
time slots for some constant $c_{2}$. During this phase, the links
transmit on the channel they were allocated in the auction phase.
If the exploration phase provided an accurate enough estimation of
the QoS and the CSMA back-off time uses enough bits for quantization,
then this phase adds no regret to the expected total regret since
the links use the optimal allocation.
\end{enumerate}
The fact that the exploitation phase takes an exponential number of
time slots does not mean it takes a long time in practice. In fact,
it only means that the lengths of the exploration and auction phases
are much shorter. Note that $T$ is finite and can be limited by the
designer, so even the last (longest) exploitation phase can still
consist of just a couple of thousands of OFDM symbols, which amounts
to only a few milliseconds. From a practical point of view, this is
the desirable packet structure since the actual transmission takes
the vast majority of the OFDM symbols while the equivalents of the
synchronization header do not cause a significant overhead. The overhead
caused by the exploration and auction phases is naturally measured
by the sum of regrets as in \eqref{eq:4}. The structure of the $k$-th
packet of our algorithm is depicted in Fig. \ref{fig:Packet}.

Our main Theorem is formulated as follows.
\begin{thm}[Main Theorem]
\label{Main Theorem}Assume that the instantaneous QoS $\left\{ q_{n,i}\left(t\right)\right\} _{t}$
are independent in $n$ and i.i.d. in time $t$, with expectations
$Q_{n,i}\in\left\{ Q_{1},\ldots,Q_{M}\right\} $ such that $Q_{i}=l_{i}\Delta_{\min}$
for a non-negative integer $l_{i}$ and a positive $\Delta_{\min}$,
and $Q_{1}<\ldots<Q_{M}$. Denote $\Delta_{\max}=Q_{M}-Q_{1}$. Let
each link run Algorithm 1 with $\varepsilon<\frac{\Delta_{\min}}{4K}$
and an exploration phase length of
\begin{equation}
c_{1}\geq K\max\left\{ \frac{81}{2}K,\frac{128}{9}\left(\frac{\Delta_{\max}}{\Delta_{\min}}\right)^{2}N^{2}\right\} \label{eq:6}
\end{equation}
Then, the expected sum of regrets is $\bar{R}\sim O\left(\log T\right)$.
\end{thm}
\begin{IEEEproof}
Lemma \ref{lem:AuctionConvergence} in Section \ref{sec:Auction Phase},
proved in Appendix D, shows that if the exploration phase succeeds
and enough bits are used for the CSMA back-off quantization, then
the exploitation phase contributes no regret to the sum of regret.
Lemma \ref{lem: exploration} in Section \ref{sec:Pure-Exploration-Phase},
proved in Appendix C, bounds from above the error probability of the
exploration phase, showing that it decreases exponentially with $k$.
The proof follows by bounding from above the expected regret using
these two facts. For details see Appendix A.
\end{IEEEproof}

\subsection{Implementation Issues}

\begin{figure*}[t]
~~~~~~~~~~~~\includegraphics[width=16cm,height=6cm]{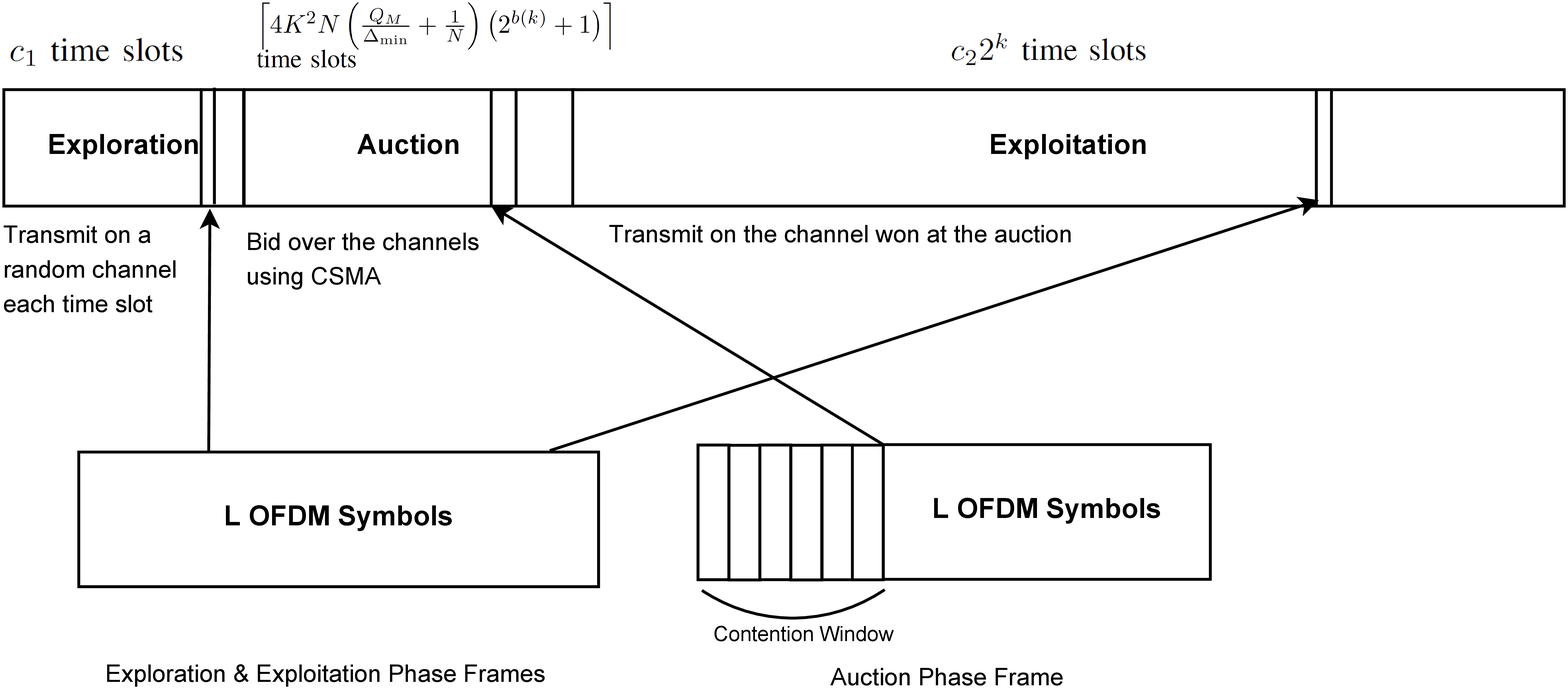}

\caption{\label{fig:Packet}The $k$-th packet of Algorithm 1}
\end{figure*}

In the problem formulation, the length of the time slots is not specified.
This is done in order to keep the theoretical framework identical
to other multi-armed bandits algorithms and measure the regret using
the same scale. However, when implementing Algorithm 1 in practice,
there is no need to assume that all time slots are of equal length.
In particular, the time slots used to implement the CSMA back-off
time can be much shorter than time slots that are used to transmit
a frame of $L$ OFDM symbols. The result, depicted in Fig. \ref{fig:Packet},
is the well-known structure of a CSMA frame, like that used in WiFi.
At the beginning of the $k$-th frame, a contention window of $2^{b\left(k\right)}$
short slots is used, followed by the transmission in the chosen channel,
over $L$ OFDM symbols. During the exploration and exploitation phases,
no contention window is needed, which makes the overhead of the contention
window negligible compared to $T$.

We also note that the computational complexity of running Algorithm
1 for each link is $O\left(K\right)$, since maximization over a $K$-sized
vectors is required.

\begin{algorithm}[t]
\caption{Distributed Channel Allocation}

\textbf{Initialization} Choose $\varepsilon<\frac{\Delta_{\min}}{4K}$.
Set $V_{n,i}\left(0\right)=0$ and $s_{n,i}\left(0\right)=0$ for
all $i$ and $b\left(0\right)=8$.
\begin{enumerate}
\item \textbf{Dither Values }- Generate $u_{n,i}$ for each $i$, independently
and uniformly at random on $\left[-\frac{\Delta_{\min}}{8N},\frac{\Delta_{\min}}{8N}\right]$.
\end{enumerate}
\textbf{For $t=1,\ldots,T$} do

\textbf{A. Exploration Phase} - For the next $c_{1}$ time slots
\begin{enumerate}
\item Choose a channel $i\in\left[1,..,K\right]$ uniformly at random.
\item Receive the reward $r_{n,i}\left(t\right)$. Updates $V_{n,i}\left(t\right)=V_{n,i}\left(t-1\right)+\eta_{i}\left(t\right)$
and $s_{n,i}\left(t\right)=s_{n,i}\left(t-1\right)+r_{n,i}\left(t\right)$.
\item Create a dithered estimation of $Q_{n,i}$ by computing $Q_{n,i}^{k}=\frac{s_{i}(t)}{V_{n,i}\left(t\right)}+u_{n,i}$
for $i=1,\ldots,K$.
\end{enumerate}
\textbf{B. Auction Phase} - set state to \textit{unassigned} and $B_{n,i}=0,\forall i$.

For the next $\left\lceil 4K^{2}N\left(\frac{Q_{M}}{\Delta_{\min}}+\frac{1}{N}\right)\left(2^{b\left(k\right)}+1\right)\right\rceil $
time slots

\textbf{Each} auction iteration\textbf{ do}
\begin{enumerate}
\item If \textit{unassigned} then
\begin{enumerate}
\item Calculate its own maximum profit:
\[
\gamma_{n}=\max_{i}\left(Q_{n,i}^{k}-B_{n,i}\right)
\]
\item Calculate its own second maximum profit:
\[
\tilde{i}_{n}=\arg\max_{k}\left(Q_{n,i}^{k}-B_{n,i}\right)
\]
\[
w_{n}=\max_{i\neq\tilde{i}_{n}}\left(Q_{n,i}^{k}-B_{n,i}\right)
\]
\item Update the price of its best channel $\tilde{i}_{n}$:
\[
B_{n,\tilde{i}_{n}}=B_{n,\tilde{i}_{n}}+\gamma_{n}-w_{n}+\varepsilon
\]
\end{enumerate}
\item During the next $2^{b\left(k\right)}$ time slots - start transmitting
in channel $\tilde{i}_{n}$ after a back-off time of
\[
\tau_{n}=f_{b\left(k\right)}\left(B_{n,\tilde{i}_{n}}\right)
\]
time slots, where $f_{b\left(k\right)}$ is a quantization of a decreasing
function, using $b\left(k\right)$ bits, such that $0\leq\tau_{n}\leq2^{b\left(k\right)}$.
\begin{enumerate}
\item Set state to \textit{assigned} if accessed channel $\tilde{i}_{n}$
before all other links and to \textit{unassigned }otherwise\textit{.}
\end{enumerate}
\item \textbf{Collision Resolution} - In the $\tau_{\max}=2^{b\left(k\right)}+1$
time slot
\begin{enumerate}
\item Transmit on channel 1 if it was assigned a channel with a collision.
\item If links sense a transmission on channel 1, then they update $b\left(k+1\right)=b\left(k\right)+1$.
\end{enumerate}
\end{enumerate}
\textbf{End}

\textbf{C. Exploitation Phase} - for the next $c_{2}2^{k}$ time slots
\begin{enumerate}
\item Transmit on the channel that it was assigned in the previous Auction
phase.
\end{enumerate}
\textbf{End}
\end{algorithm}

\section{Exploration Phase - Estimation of the QoS\label{sec:Pure-Exploration-Phase}}

In this section, we analyze the performance of the exploration phase
and its contribution to the expected sum-regret. The distributed algorithm
of \cite{naparstek2013optimal} assumes each link knows its CSI, or
the possible QoS each channel supports. Our algorithm lifts this assumption
by working on online estimations of the CSI (or QoS) instead. Each
link obtains these estimations by randomly exploring the different
$K$ channels and averaging the instantaneous measurements of the
QoS of each channel.

The exploration phase does not require the links to know the total
number of links $N$ or the total duration of transmission $T$. Hence,
links cannot use a single long enough exploration phase at the beginning,
since they want the exploration error probability to be designed according
to $T$ and $N$. The packet structure in Fig. \ref{fig:Packet} maintains
the required balance. In each packet, only a constant number $c_{1}$
of time slots is dedicated to exploration, but the estimation of the
$k$-th exploration phase uses all the previous exploration phases.

The estimated QoS of the channels is needed for the next auction phase
to converge to the optimal allocation. However, due to its distributed
nature, ties cannot be arbitrarily broken. Hence, the exploration
phase needs to output accurate enough estimates that guarantee that
there will be no ties in the bids in the auction algorithm. For that
purpose, after the estimation of the expected QoS is completed, artificial
dither noise is added to the estimated values. This dither values
are generated in advance independently and uniformly at random on
a small interval. The following lemma characterizes the required estimation
accuracy of the exploration phase, taking into account the dither
noise.
\begin{lem}
\label{lem:Percision}Denote the dithered estimations of the expected
QoS values in packet $k$ by $\left\{ Q_{n,i}^{k}\right\} $. Assume
that $\left|Q_{n,i}^{k}-Q_{n,i}-u_{n,i}\right|\leq\Delta$ for each
link $n$ and channel $i$ for some positive $\Delta$. If $\Delta<\frac{3\Delta_{\min}}{8N}$
then
\begin{equation}
\arg\underset{a_{1},...,a_{N}}{\max}\sum_{n=1}^{N}Q_{n,a\left(n\right)}=\arg\underset{a_{1},...,a_{N}}{\max}\sum_{n=1}^{N}Q_{n,a\left(n\right)}^{k}.\label{eq:10}
\end{equation}
\end{lem}
\begin{IEEEproof}
The proof follows from the fact that if $Q_{n,i}^{k}$ and $Q_{n,i}$
are close enough for every $i$ and $n$, then the optimal assignment
on $\left\{ Q_{n,i}^{k}\right\} $ and $\left\{ Q_{n,i}\right\} $
must be identical. For details see Appendix B.
\end{IEEEproof}
The following lemma concludes this section by providing an upper bound
for the probability that the estimation for packet $k$ failed. The
fact that this error probability exponentially vanishes with $k$,
allows us to limit the number of exploration time slots to $c_{1}$,
keeping the overhead caused by the exploration phase negligible.
\begin{lem}[Exploration Error Probability]
\label{lem: exploration} Denote the dithered estimations of the
expected QoS values in packet $k$ by $\left\{ Q_{n,i}^{k}\right\} $.
If the length of the exploration phase satisfies $c_{1}\geq K\max\left\{ \frac{81}{2}K,\frac{128}{9}\left(\frac{\Delta_{\max}}{\Delta_{\min}}\right)^{2}N^{2}\right\} $,
then after the $k$-th packet we have
\begin{equation}
P_{e,k}\triangleq\Pr\left(\underset{n,i}{\max}\left|Q_{n,i}^{k}-Q_{n,i}\right|>\frac{3\Delta_{\min}}{8N}\right)\leq3NKe^{-k}.\label{eq:11}
\end{equation}
\end{lem}
\begin{IEEEproof}
The proof uses Hoeffding's bound on both $\left|Q_{n,i}^{k}-Q_{n,i}\right|$
and the number of samples of $Q_{n,i}$ without collision. For details
see Appendix C.
\end{IEEEproof}

\section{Auction Phase - Converging to the Optimal Allocation\label{sec:Auction Phase}}

We adopt the distributed auction in \cite{naparstek2013optimal} as
the basis for our auction phase. The multi-armed bandit problem uses
a discrete time axis. Hence, a continuous back-off time as used in
\cite{naparstek2013optimal} is not possible. From a practical perspective,
links cannot implement a truly continuous delay but a quantized one.
With integer quantized delays, it is possible that two links use the
same delay for the same channel although their continuous bids are
different. In this case, they cannot agree on which of them won the
bid and got the channel. It is clear that for a fine enough quantization,
these bidding collisions will be avoided. However, due to the distributed
nature of the problem, links do not know in advance what is considered
a fine enough quantization. We propose a collision resolution algorithm
that increases the quantization bits, described in step 3 in the Auction
phase in Algorithm 1. Links coordinate their quantization by employing
a ``voting turn'' that only uses the fact that all links can sense
a single channel of their choice. In this special time slot, links
listen to channel 1 which is used to signal if a collision occurred
for some of the links.

Another issue to be resolved is where the continuous bids of two links
$m$ and $n$ are identical, $B_{n,i}=B_{m,j}$. Since there is no
auctioneer, the links cannot agree on an arbitrary tie braking without
communication. Hence, identical bids can prevent the CSMA auction
algorithm from converging to the optimal solution. In order to avoid
this problem, the auction phase uses a noisy version of the estimated
expected rewards from the exploration phase. This noise is an artificial
dither added by the links independently such that the probability
for identical bids will be zero.
\begin{lem}
\label{lem:Identical Bids}After the $k$-th exploration phase we
have $\Pr\left(B_{n,i}=B_{m,j}\right)=0$ for any $n\neq m$ and any
$i,j$.
\end{lem}
\begin{IEEEproof}
Due to the continuous (uniform) distribution of $u_{n,i}$ and $u_{m,j}$,
for any $m\neq n$ and $i,j$, the probability that $Q_{n,i}^{k}=\frac{s_{i}(t)}{o_{i}}+u_{n,i}=Q_{m,j}^{k}=\frac{s_{j}(t)}{o_{j}}+u_{m,j}$
is zero. Since any bid $B_{n,i}$ is a linear combination of rewards
and $\varepsilon$, also the probability that at a certain iteration
of the auction algorithm $B_{n,i}=B_{m,j}$ is zero.
\end{IEEEproof}
We emphasize that Lemma 6, and Lemma 7 below, only help to show (in
Lemma 8) that Algorithm 1 eventually converges to the optimal solution.
Links start transmitting data from the first packet, using a possibly
suboptimal allocation in the exploitation phase. Hence, Algorithm
1 is likely to perform well much before convergence to the optimal
allocation occurred. Nevertheless, our simulations in Section VI suggest
that convergence to the optimal allocation occurs very fast, already
in the first or the second packet.
\begin{lem}
Algorithm 1 converges to some final value $b_{f}$, i.e., there exists
a $k_{0}$ such that $b(k)=b_{f}$ for all $k>k_{0}$.
\end{lem}
\begin{IEEEproof}
Consider two different bids $B_{n,i}\neq B_{m,j}$ of two different
links $n\neq m$, and assume that after quantization to $b\left(k\right)$
bits we have $f_{b\left(k\right)}\left(B_{n,i}\right)=f_{b\left(k\right)}\left(B_{m,j}\right)$.
In this case, the links will detect a collision after the auction
phase and will increase the number of bits used for quantization.
Since $B_{n,i}-B_{m,j}$ is a sum of rewards and some multiplication
of $\varepsilon$, for large enough $b\left(k\right)=b^{*}$, we have
$f\left(B_{n,i}\right)\neq f\left(B_{m,j}\right)$ for any $m,n,i,j$
such that $n\neq m$ and $B_{n,i}\neq B_{m,j}$. Hence, $b\left(k\right)$
will not increase above $b^{*}$, since collisions between $B_{n,i}\neq B_{m,j}$
cannot occur with $b\left(k\right)=b^{*}$. Collisions from identical
bids $B_{n,i}=B_{m,j}$ do not occur simply because their probability
is zero, as shown in Lemma \ref{lem:Identical Bids}.
\end{IEEEproof}
\begin{lem}
\label{lem:AuctionConvergence} Assume that $b(k')=b_{f}$ for all
$k'>k$. If the $k$-th exploration phase succeeded, then the $k$-th
auction phase converges to an allocation $a_{1},\ldots,a_{N}$ such
that $\left|\sum_{n=1}^{N}Q_{n,a_{n}}^{k}-\underset{a_{1},...,a_{N}}{\max}\sum_{n=1}^{N}Q_{n,a_{n}}^{k}\right|\leq\varepsilon$
in less than $\frac{KN}{\varepsilon_{k}}\left(Q_{M}+\frac{\Delta_{\min}}{8N}\right)2^{b\left(k\right)}$
time slots with probability 1. If $\varepsilon<\frac{3\Delta_{\min}}{4K}$,
then the auction phase converges to $\arg\underset{a}{\max}\sum_{n=1}^{N}Q_{n,a_{n}}$.
\end{lem}
\begin{IEEEproof}
The proof follows from the convergence and performance guarantees
proven in \cite{Naparstek2013} together with Lemma \ref{lem: exploration}.
For details see Appendix D.
\end{IEEEproof}

\section{Simulation Results}

In this section, we demonstrate the performance of Algorithm 1 using
computer simulations. We compared Algorithm 1 with the centralized
Hungarian method, random channel selection and the E3 algorithm in
\cite{Nayyar2016}. The Hungarian method requires some central entity
to know the CSI of all users. Requiring much less information, the
E3 algorithm assumes that each user can decode which channel each
of the other users has chosen. Our algorithm requires even much less
information - each user only needs to sense if there is a transmission
on a given channel. The role of the simulations of this section is
to show that despite our much stricter information constraints, our
algorithm performs almost exactly well as the E3 algorithm and even
the optimal Hungarian algorithm. The comparison with the random channel
selection assures that an algorithm that does not strive to converge
to the optimal allocation performs very badly. This serves to show
that the problem is far from being degenerated or trivial.

We verified our algorithm under various network scenarios consisting
of different path losses and fading environments. The channel was
divided into $N$ sub-channels and we used $N=K=10$. The transmit
power spectral density (PSD) was fixed at $12\mbox{dBm}$ for each
user. The users were assumed to be moving at a speed of $3\mbox{km/h}$.
We used a transmission duration of $T=10^{5}$ time slots, with a
single OFDM symbol per time slot ($L=1$). Our transmission packet
(see Fig. \ref{fig:Packet}) has an exploration phase of 800 OFDM
symbols and an auction phase of 500 OFDM symbols. Each experiment
consists of averaging 100 independent realizations.

\begin{figure*}[t]
\begin{minipage}[t]{0.5\columnwidth}%
\includegraphics[width=9cm,height=5cm]{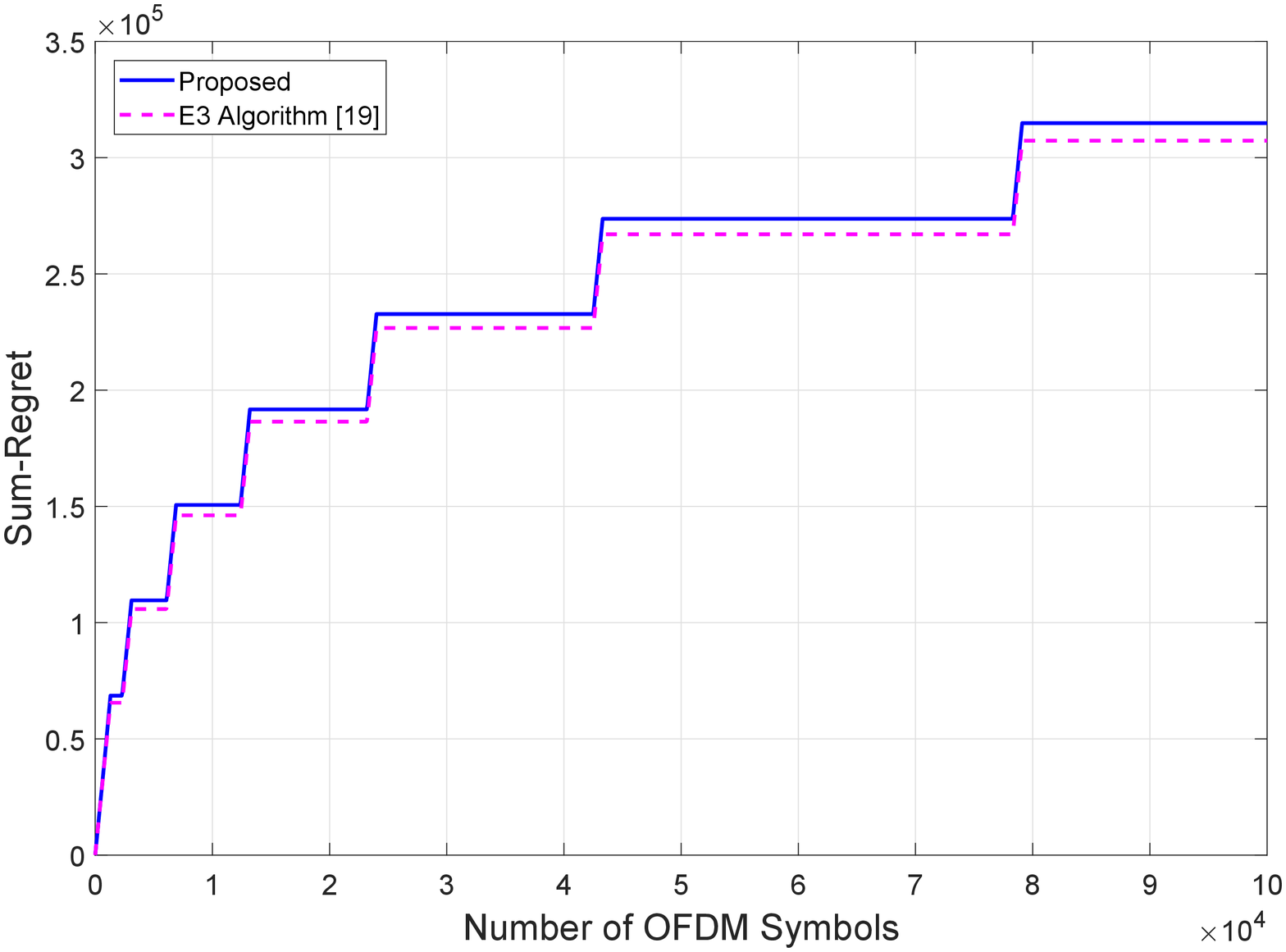}%
\end{minipage}%
\begin{minipage}[t]{0.5\columnwidth}%
\includegraphics[width=9cm,height=5cm]{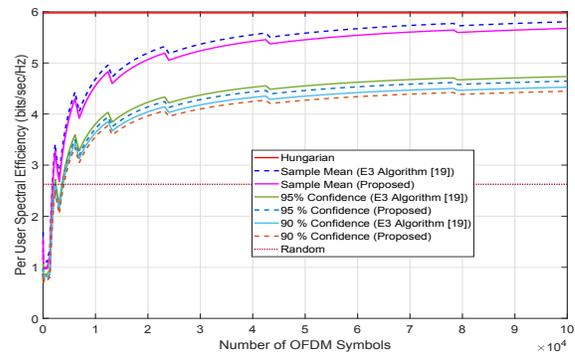}%
\end{minipage}

\caption{\label{fig:Rayliegh}Performance evaluation over i.i.d. Rayleigh fading
channel. Simulation parameters are: N = K = 10, explore length= 800
OFDM symbols, and auction length = 500 OFDM symbols.}
\end{figure*}

\begin{figure*}[t]
\begin{minipage}[t]{0.5\columnwidth}%
\includegraphics[width=9cm,height=6cm]{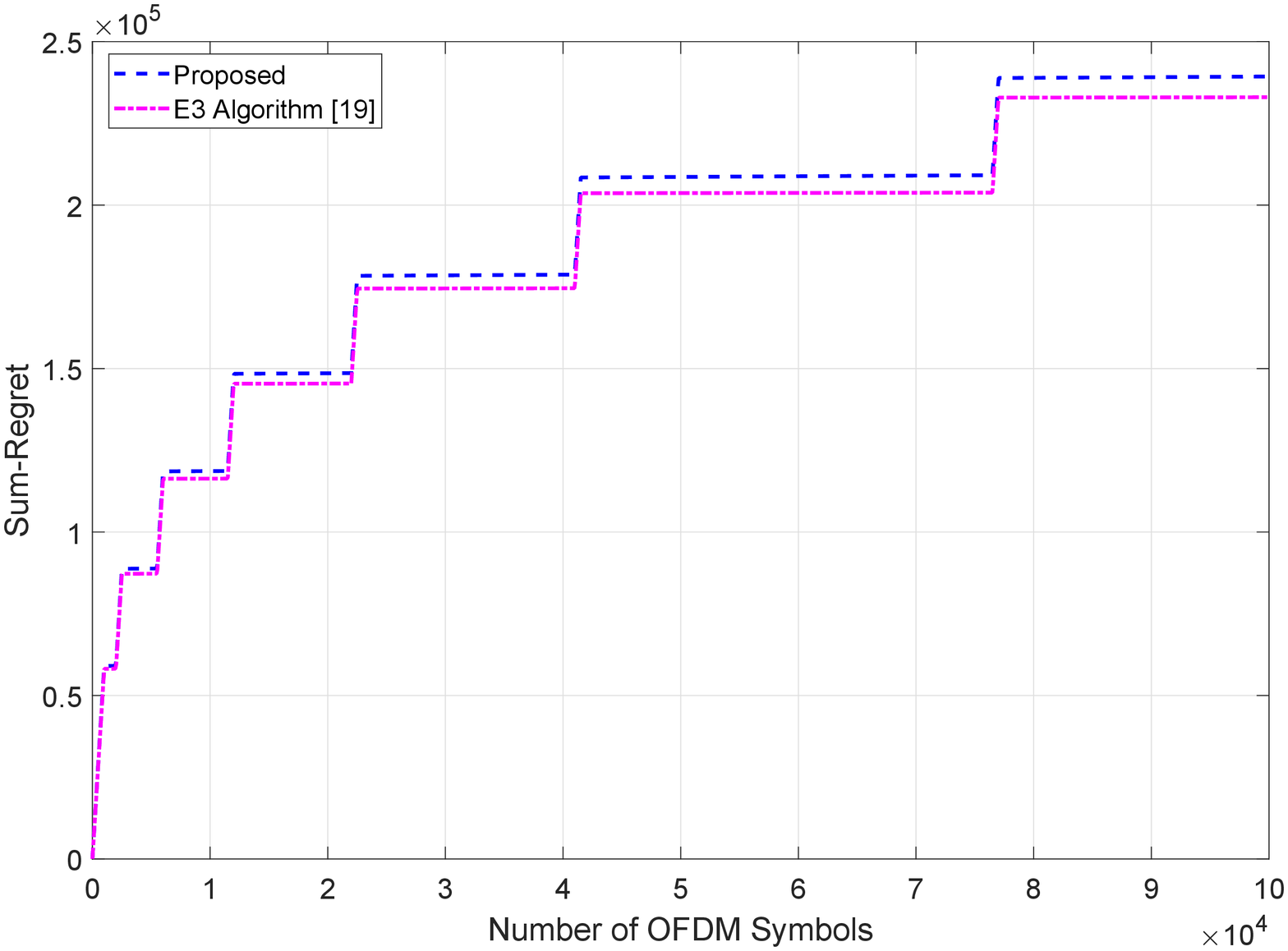}%
\end{minipage}%
\begin{minipage}[t]{0.5\columnwidth}%
\includegraphics[width=9cm,height=6cm]{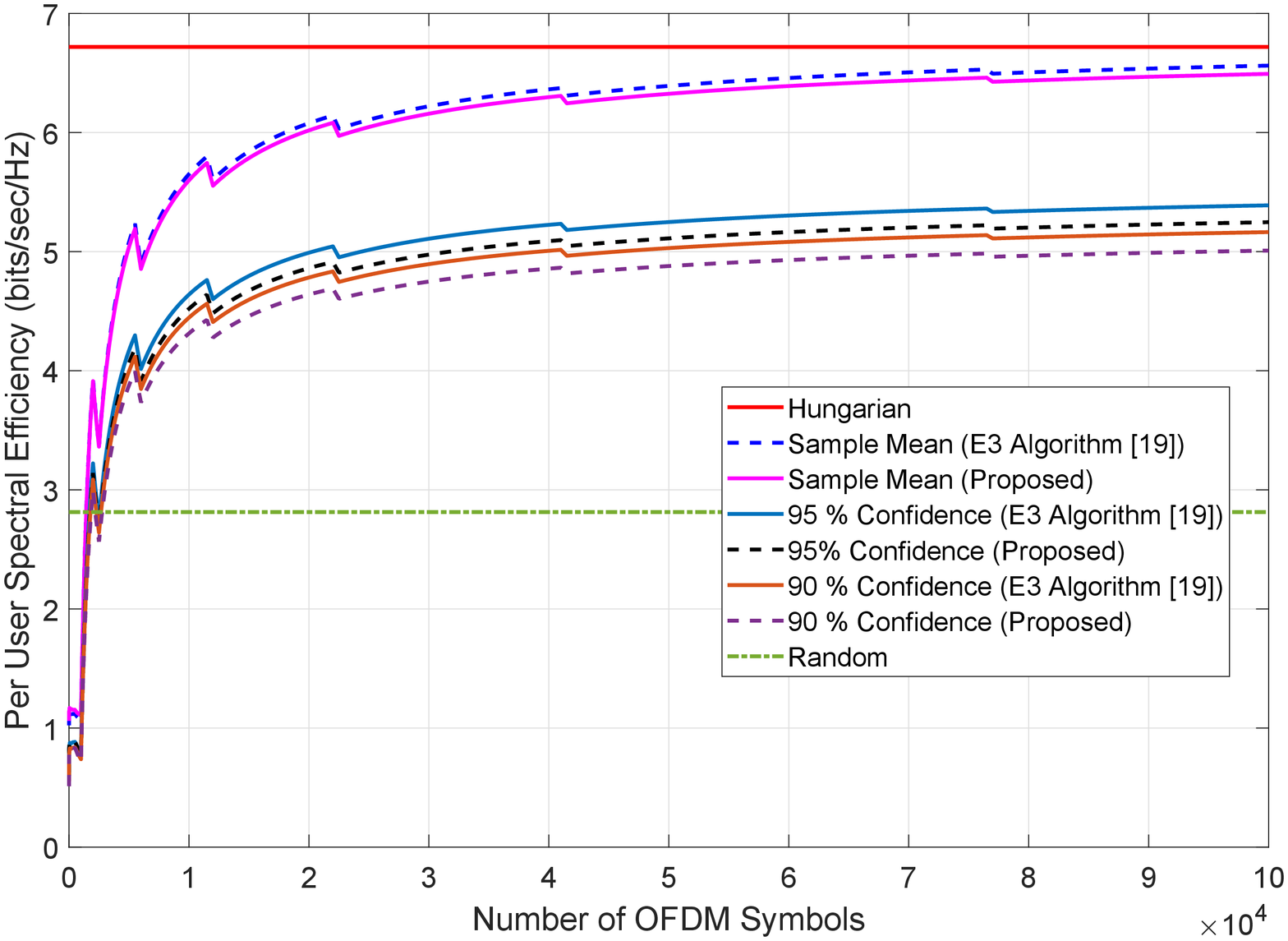}%
\end{minipage}

\caption{\label{fig:LTEnoInterference}Performance evaluation over LTE fading
channel. Simulation parameters are: $N=K=10$, explore length= 800
OFDM symbols, and auction length = 500 OFDM symbols.}
\end{figure*}

\begin{figure*}[t]
\begin{minipage}[t]{0.5\columnwidth}%
\includegraphics[width=9cm,height=5cm]{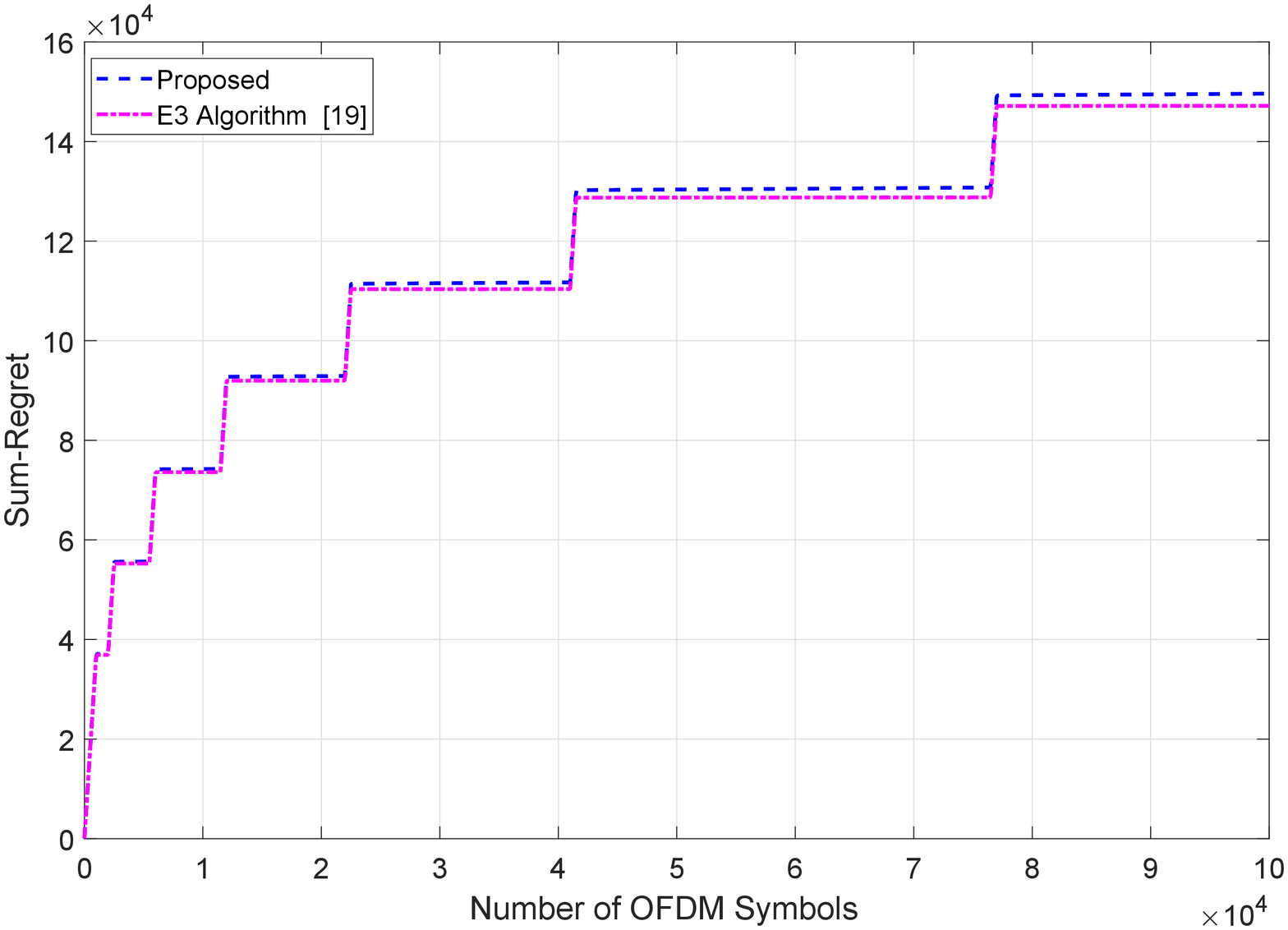}%
\end{minipage}%
\begin{minipage}[t]{0.5\columnwidth}%
\includegraphics[width=9cm,height=5cm]{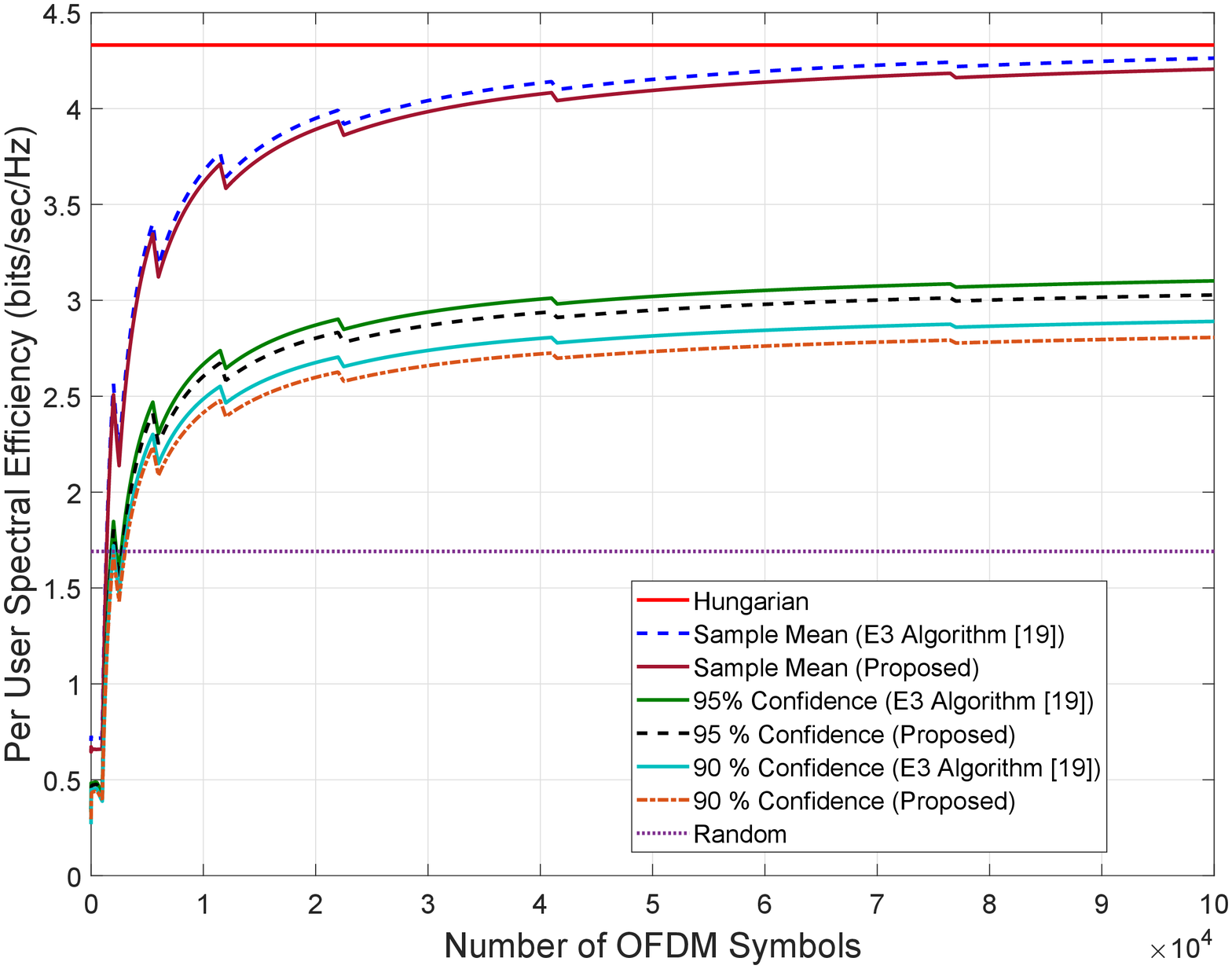}%
\end{minipage}

\caption{\label{fig:LTE}Performance evaluation over LTE fading channel with
alien interference. Simulation parameters: $N=K=10$, explore length=
500 OFDM symbols, and auction length = 500 OFDM symbols.}
\end{figure*}

First, we considered an ad-hoc network of $N$ links that are uniformly
distributed on disk with a radius of 500 m. The central carrier frequency
was 2 $\mbox{GHz}$ with a per-user transmission bandwidth of 200
$\mbox{KHz}$. The path loss was computed using path loss exponent
of $\alpha=4$. We considered two types of channel models: i.i.d.
Rayleigh fading channel and the extended pedestrian A model (EPA)
of the LTE standard with 9 random taps. In Fig. \ref{fig:Rayliegh}
the sum-regret of our algorithm is compared to that of the E3 algorithm
\cite{Nayyar2016} under an i.i.d. Rayleigh fading channel. It is
evident that the performance of both algorithms is essentially identical,
despite the fact that our algorithm uses no communication between
users as the E3 algorithm \cite{Nayyar2016} does. Both algorithms
have an expected sum-regret that increases like $\log T$ and both
converge to the optimal allocation already at the first packets. In
Fig. \ref{fig:LTE}, we present the spectral efficiency performance
of both algorithms together with the confidence intervals of 90\%
and 95\%, where again all performances are very similar between our
algorithm and the E3 algorithm \cite{Nayyar2016}. It also shows that
the proposed algorithm approaches the optimal performance within a
few packets, does much better than a random selection and behaves
very similarly in all realizations. We have repeated the above experiment
for the more realistic scenario of LTE channels. Fig. \ref{fig:LTEnoInterference}
again confirms that our performance is identical to that of the E3
algorithm \cite{Nayyar2016}.

Next, in Fig. \ref{fig:LTE} we demonstrate the performance of the
proposed algorithm in the presence of alien interference for LTE channels.
In this scenario, we considered four interferers that use four out
of $K=10$ available channels. These interfering nodes are randomly
located outside the network disk and within a distance of 500 m from
the annular region of the disk. It can be seen from the right graph
in Fig. \ref{fig:LTE} that the spectral efficiency is reduced by
\textasciitilde 2 bits/sec/Hz. However, the proposed algorithm achieves
the optimal performance within few thousand symbols similar to the
interference-free case, as shown in Fig. \ref{fig:LTEnoInterference}.
This scenario again confirms that our performance is identical to
that of the E3 algorithm \cite{Nayyar2016}.

\begin{figure*}[t]
\begin{minipage}[t]{0.5\columnwidth}%
\includegraphics[width=9cm,height=6cm]{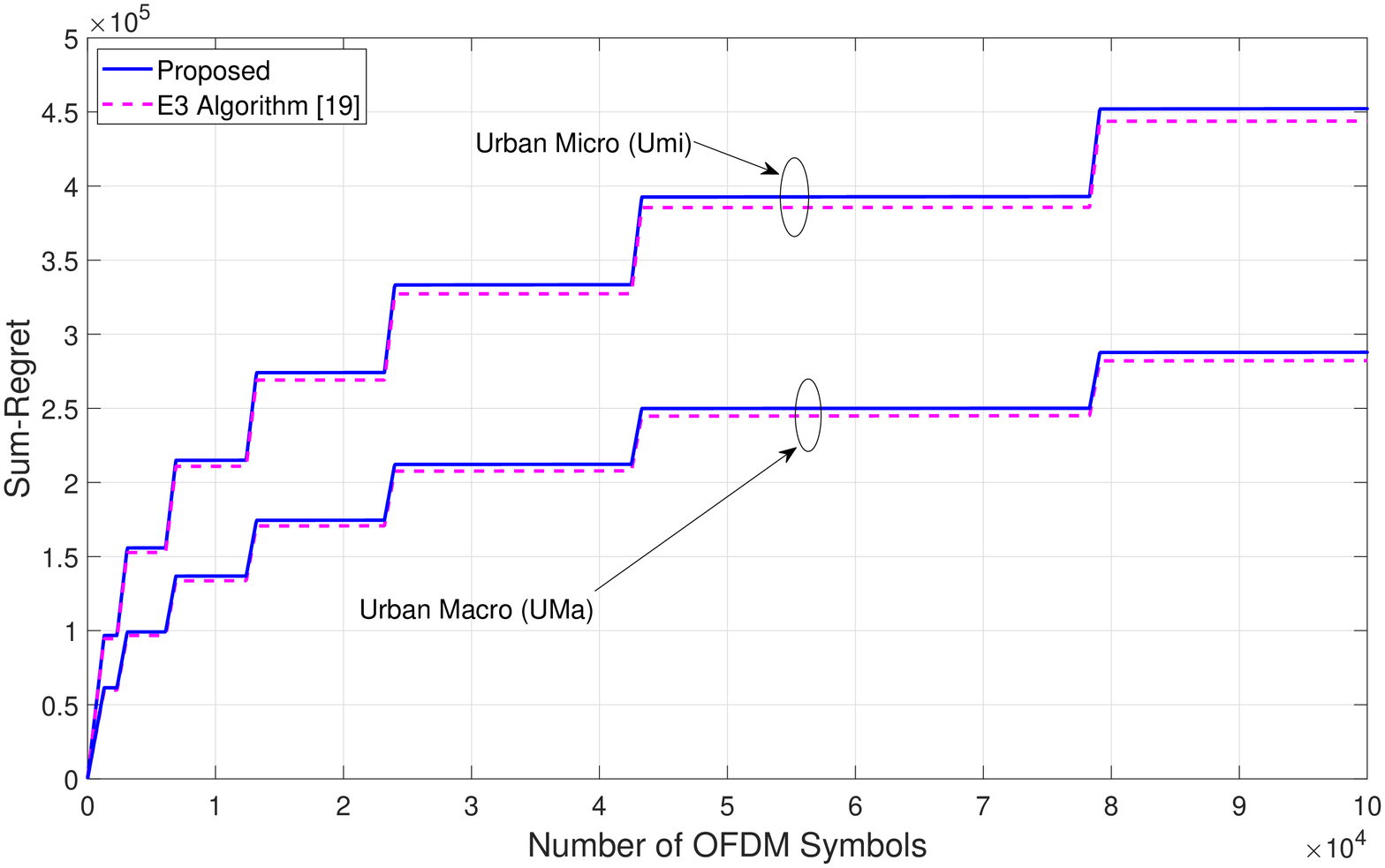}%
\end{minipage}%
\begin{minipage}[t]{0.5\columnwidth}%
\includegraphics[width=9cm,height=6cm]{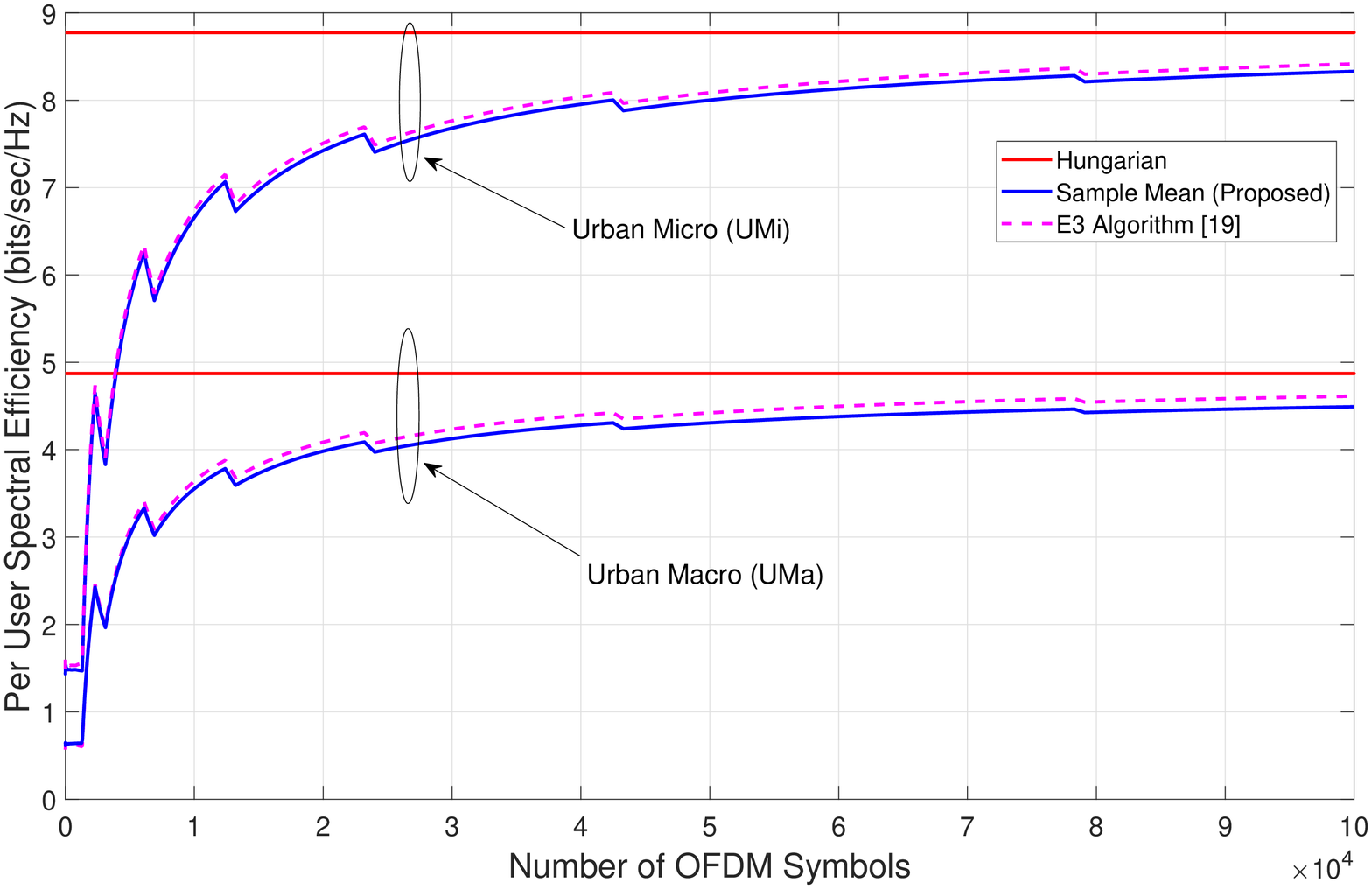}%
\end{minipage}

\caption{\label{fig:5G}Performance evaluation over 5G fading channels with
different path loss models. Simulation parameters: $N=K=10$, explore
length= 800 OFDM symbols, and auction length = 500 OFDM symbols.}
\end{figure*}

Finally, we considered a 5G system with more realistic channel scenarios
consisting of pathloss, short-term fading, and long-term shadowing.
We computed the path loss from empirical models of urban macro (UMa)
in the distance range of $45\mbox{m}$ to $1429\mbox{m}$ and urban
micro-street canyon (UMi-SC) in the distance range of $19\mbox{m}$
to $272\mbox{m}$ \cite{Meredith2016,Sun2016}. The shadowing factor
is $6\mbox{dB}$ and $7.8\mbox{dB}$ for the UMa and UMi-SC models,
respectively. The fading channel consists of tapped delay line (TDL-A)
model with 23 taps. The central carrier frequency was $6\mbox{GHz}$
with a per-user transmission bandwidth of $720\mbox{KHz}$. The results
in Fig. \ref{fig:5G} demonstrate that the all the realistic channel
phenomena we simulated do not prevent the proposed algorithm from
quickly converging to the optimal solution.

The simulations in this section provide an additional solid support
that our algorithm offers the same performance as \cite{Nayyar2016}
but with significantly less requirements from the devices. Specifically,
we require no information exchange between different links as required
in \cite{Nayyar2016} and we only use the sensing of a single channel
each time slot instead of sensing all channels simultaneously.

\section{Conclusions\label{sect:conclude}}

In this paper, we suggested a distributed algorithm for channel allocation
with time varying-channels where links initially have no estimation
for the statistics of the channels. Learning the statistics of the
channels in real-time (exploration) comes at the expanse of using
the best known channels (exploitation). The scenario is described
by a multi-armed bandit game where a collision occurs if two or more
links transmit on the same channel. We proved that our algorithm achieves
the optimal order of regret - $O\left(\log T\right)$. Our algorithm
is based on a distributed auction algorithm that uses CSMA to avoid
the need for an auctioneer (base station). In contrast to the state-of-the-art
algorithms, our algorithm requires neither centralized management
nor any communication between different links, which makes it very
relevant to cognitive ad-hoc networks. Our algorithm only requires
sensing a single channel at each time slot, which is $K$ times less
than the state-of-the-art algorithms, where $K$ is the number of
channels. Only a detection of whether there are transmissions on this
channel is required, and no decoding and demixing operations are needed
to discern which user chose which channel. From a practical point
of view, this results in a significant complexity reduction of the
physical layer design. Simulations show that our algorithm performs
very well on realistic LTE and 5G channels.

\appendices{}

\section{Proof of Theorem \ref{Main Theorem}}
\begin{IEEEproof}
Denote the number of packets that start within $T$ time slots by
$E$. Let $k_{0}$ be the index of a sufficiently large packet. We
compute the expected total regret as follows:
\begin{equation}
\bar{R}\leq\underbrace{\sum_{k=1}^{k_{0}}\bar{R}_{k}}_{\bar{R}_{0}}+\sum_{k=k_{0}+1}^{E}\bar{R}_{k}\label{eq:7}
\end{equation}
where $\bar{R}_{k}$ is the expected total regret of packet $k$ and
$\bar{R}_{0}$ is a constant with respect to $T$. Denote by $P_{e,k}$
the error probability of the exploration of packet $k$. In Lemma
\ref{lem:AuctionConvergence}, we prove that if the exploration phase
succeeded and the number of quantization bits $b\left(k\right)$ for
the CSMA delay is large enough, then the auction phase is guaranteed
to converge to the optimal solution of $\eqref{eq:5}$ for any $\varepsilon<\frac{\Delta_{\min}}{4K}$.
This optimal allocation is played in the exploitation phase, which
adds no additional regret to the total regret. We prove in Lemma \ref{lem: exploration}
that if \eqref{eq:6} holds then, $P_{e,k}\leq3NKe^{-k}$. Hence,
we obtain for a large enough $k$ such that $b\left(k\right)$ is
sufficiently large that

\begin{multline}
\bar{R}_{k}\leq\left(c_{1}+4K^{2}N\left(\frac{Q_{M}}{\Delta_{\min}}+\frac{1}{N}\right)\left(2^{b\left(k\right)}+1\right)+1\right)N+3NKc_{2}\left(\frac{2}{e}\right)^{k}N\leq\\
2\left(c_{1}+4K^{2}N\left(\frac{Q_{M}}{\Delta_{\min}}+\frac{1}{N}\right)\left(2^{b_{f}}+1\right)\right)N\label{eq:8}
\end{multline}
for some constant $b_{f}$. We conclude that
\begin{multline}
\bar{R}\leq\bar{R}_{0}+\sum_{k=k_{0}+1}^{E}\bar{R}_{k}\underset{(a)}{\leq}\bar{R}_{0}+2\left(c_{1}+4K^{2}N\left(\frac{Q_{M}}{\Delta_{\min}}+\frac{1}{N}\right)\left(2^{b_{f}}+1\right)\right)NE\underset{(b)}{\leq}\\
\bar{R}_{0}+2\left(c_{1}+4K^{2}N\left(\frac{Q_{M}}{\Delta_{\min}}+\frac{1}{N}\right)\left(2^{b_{f}}+1\right)\right)N\log_{2}\left(\frac{T}{c_{2}}+2\right)\label{eq:9}
\end{multline}
where in (a) we used the fact that completing the last packet to be
a full packet only increases $\bar{R}_{k}$. In (b) we used $T>\sum_{k=1}^{E-1}c_{2}2^{k}\geq c_{2}\left(2^{E}-2\right)$,
which yields $E\leq\log_{2}\left(\frac{T}{c_{2}}+2\right)$.
\end{IEEEproof}

\section{Proof of Lemma \ref{lem:Percision}}
\begin{IEEEproof}
Recall that $\Delta_{\min}=\underset{i\neq j}{\min}\left|Q_{i}-Q_{j}\right|$.
For all $n$ and $i$ we have $Q_{n,i}^{k}=Q_{n,i}+z_{n,i}+u_{n,i}$
such that $\left|u_{n,i}\right|\leq\frac{\Delta_{\min}}{8N}$, and
we assume that $\left|z_{n,i}\right|\leq\Delta$. In the perturbed
assignment problem, an optimal assignment $a^{1}\in\arg\underset{a_{1},...,a_{N}}{\max}\sum_{n=1}^{N}Q_{n,a\left(n\right)}$
performs at least as well as
\begin{equation}
\sum_{n=1}^{N}Q_{n,a^{1}\left(n\right)}^{k}=\sum_{n=1}^{N}\left(Q_{n,a^{1}\left(n\right)}+z_{n,a^{1}\left(n\right)}+u_{n,a^{1}\left(n\right)}\right)\geq\sum_{n=1}^{N}Q_{n,a^{1}\left(n\right)}-\left(\Delta+\frac{\Delta_{\min}}{8N}\right)N.\label{eq:16}
\end{equation}
Any non optimal assignment $a$ performs at most as well as
\begin{equation}
\sum_{n=1}^{N}Q_{n,a\left(n\right)}^{k}\leq\sum_{n=1}^{N}\left(Q_{n,a^{2}\left(n\right)}+z_{n,a\left(n\right)}+u_{n,a\left(n\right)}\right)\leq\sum_{n=1}^{N}Q_{n,a^{2}\left(n\right)}+\left(\Delta+\frac{\Delta_{\min}}{8N}\right)N\label{eq:17}
\end{equation}
where $a^{2}$ is an assignment with the second best objective. For
any two assignments $a\neq a'$ with a different sum of QoS we have
\begin{equation}
\sum_{n=1}^{N}Q_{n,a\left(n\right)}-\sum_{n=1}^{N}Q_{n,a'\left(n\right)}\geq\Delta_{\min}.\label{eq:18}
\end{equation}
We conclude that for any non optimal $a$
\begin{equation}
\sum_{n=1}^{N}Q_{n,a^{1}\left(n\right)}^{k}-\sum_{n=1}^{N}Q_{n,a\left(n\right)}^{k}\underset{\left(a\right)}{\geq}\left(\sum_{n=1}^{N}Q_{n,a^{1}\left(n\right)}-\sum_{n=1}^{N}Q_{n,a^{2}\left(n\right)}\right)-\left(2\Delta+\frac{\Delta_{\min}}{4N}\right)N\underset{\left(b\right)}{\geq}\frac{3\Delta_{\min}}{4}-2\Delta N\underset{\left(c\right)}{>}0\label{eq:19}
\end{equation}
where (a) follows from \eqref{eq:16} and \eqref{eq:17}, (b) from
\eqref{eq:18} and (c) holds for $\Delta<\frac{3\Delta_{\min}}{8N}$.
\end{IEEEproof}

\section{Proof of Lemma \ref{lem: exploration}}
\begin{IEEEproof}
After the $k$-th exploration phase, the number of samples that are
used for estimating the expected QoS is $T_{e}\left(k\right)=c_{1}k$.
Let $A_{n,i}\left(t\right)$ be the indicator that is equal to one
if only link $n$ chose channel $i$ at time slot $t$. Also define
$V_{n,i}\left(t\right)\triangleq\sum_{\tau}A_{n,i}\left(\tau\right)$,
which is the number of times that link $n$ has used channel $i$
with no collision, up to time slot $t$ and define $V_{\min}=\underset{n,i}{\min}V_{n,i}\left(t\right)$.
Recall that $\Delta_{\max}=Q_{M}-Q_{1}$ and define the estimation
error of channel $i$ for link $n$ by
\begin{equation}
\xi_{n,i}\triangleq\left|\frac{1}{V_{n,i}\left(t\right)}\sum A_{n,i}\left(\tau\right)r_{n,i}\left(\tau\right)-Q_{n,i}\right|.\label{eq:16-1}
\end{equation}
Denote by $E$ the event in which there exists a link $n$ that has
$\xi_{n,i}\geq\Delta$ for some channel $i$. We have
\begin{equation}
\Pr\left(E|V_{\min}=v\right)=\Pr\left(\bigcup_{i=1}^{K}\bigcup_{n=1}^{N}\left\{ \xi_{n,i}\geq\Delta\,|\,V_{\min}=v\right\} \right)\underset{\left(a\right)}{\leq}NK\underset{n,i}{\max}\Pr\left(\xi_{n,i}\geq\Delta\,|\,V_{\min}=v\right)\underset{\left(b\right)}{\leq}2NKe^{-\frac{2\Delta^{2}}{\Delta_{\max}^{2}}v}.\label{eq:20}
\end{equation}
where (a) follows by taking the union bound over all links and channels
and (b) from using Hoeffding's inequality for bounded variables \cite{Hoeffding1963}.
Since the exploration phase consists of uniform and independent arm
choices we have
\begin{equation}
\Pr\left(A_{n,i}\left(t\right)=1\right)=\frac{1}{K}\left(1-\frac{1}{K}\right)^{N-1}.\label{eq:21}
\end{equation}
Therefore
\begin{multline}
\Pr\left(V_{\min}<\frac{T_{e}\left(k\right)}{4K}\right)=\Pr\left(\bigcup_{i=1}^{K}\bigcup_{n=1}^{N}\left\{ V_{n,i}\left(t\right)\leq\frac{T_{e}\left(k\right)}{4K}\right\} \right)\underset{\left(a\right)}{\leq}NK\Pr\left(V_{1,1}\left(t\right)\leq\frac{T_{e}\left(k\right)}{4K}\right)\underset{\left(b\right)}{\leq}\\
NKe^{-2\frac{1}{K^{2}}\left(\left(1-\frac{1}{K}\right)^{N-1}-\frac{1}{4}\right)^{2}T_{e}\left(k\right)}\underset{\left(c\right)}{\leq}NKe^{-\frac{2}{81K^{2}}T_{e}\left(k\right)}\label{eq:22}
\end{multline}
where (a) follows from the union bound, (b) from Hoeffding's inequality
for Bernoulli random variables and (c) since $K\geq N$ and $\left(1-\frac{1}{K}\right)^{K-1}-\frac{1}{4}\geq e^{-1}-\frac{1}{4}>\frac{1}{9}$.
We conclude that
\begin{multline}
P_{e,k}\leq\Pr\left(E\right)=\sum_{v=0}^{T_{e}\left(k\right)}\Pr\left(E|V_{\min}=v\right)\Pr\left(V_{\min}=v\right)\leq\\
\sum_{v=0}^{\left\lfloor \frac{T_{e}\left(k\right)}{4K}\right\rfloor }\Pr\left(V_{\min}=v\right)+\sum_{\left\lceil \frac{T_{e}\left(k\right)}{4K}\right\rceil +1}^{T_{e}\left(k\right)}\Pr\left(E|V_{\min}=v\right)\Pr\left(V_{\min}=v\right)\leq\Pr\left(V_{\min}<\frac{T_{e}\left(k\right)}{4K}\right)+\Pr\left(E|\,\,V_{\min}\geq\frac{T_{e}\left(k\right)}{4K}\right)\\
\underset{\left(a\right)}{\leq}2NKe^{-\frac{\Delta^{2}c_{1}}{2K\Delta_{\max}^{2}}k}+NKe^{-\frac{2c_{1}k}{81K^{2}}}\label{eq:23}
\end{multline}
where (a) follows from \eqref{eq:20} and \eqref{eq:22}. We choose
$\Delta=\frac{3\Delta_{\min}}{8N}$ and $c_{1}=K\max\left\{ \frac{81}{2}K,\frac{128}{9}\left(\frac{\Delta_{\max}}{\Delta_{\min}}\right)^{2}N^{2}\right\} $
to obtain
\begin{equation}
P_{e,k}\leq2NKe^{-\frac{9\Delta_{\min}^{2}c_{1}}{128K\Delta_{\max}^{2}N^{2}}k}+NKe^{-\frac{2c_{1}k}{81K^{2}}}\leq3NKe^{-k}.\label{eq:24}
\end{equation}
\end{IEEEproof}

\section{Proof of Lemma \ref{lem:AuctionConvergence}}
\begin{IEEEproof}
In Lemma 3 of \cite{naparstek2013optimal} it is shown that the number
of iterations $I_{\textrm{auc}}$ of the distributed auction algorithm
with $\varepsilon$ is bounded by
\begin{equation}
I_{\textrm{auc}}\leq KN+\frac{K}{\varepsilon}\sum_{n=1}^{N}Q_{n,i}^{k}\underset{\left(a\right)}{\leq}KN+\frac{KN}{\varepsilon}\left(Q_{M}+\frac{\Delta_{\min}}{8N}\right)\label{eq:12}
\end{equation}
where (a) follows since $Q_{n,i}^{k}\leq Q_{M}+\frac{\Delta_{\min}}{8N}$
for all $n$ and $i$. Note that each iteration of the auction phase
takes $2^{b\left(k\right)}+1$ time slots. If the $k$-th exploration
phase succeeded we have $\underset{n,i}{\max}\left|Q_{n,i}^{k}-Q_{n,i}\right|<\frac{3\Delta_{\min}}{8N}$.
For any two allocations $a\neq a'$ with a different sum of QoS we
have
\begin{equation}
\left|\sum_{n=1}^{N}Q_{n,a_{n}}-\sum_{n=1}^{N}Q_{n,a_{n}'}\right|\geq\Delta_{\min}\label{eq:23-1}
\end{equation}
Hence
\begin{multline}
\left|\sum_{n=1}^{N}Q_{n,a_{n}}^{k}-\sum_{n=1}^{N}Q_{n,a_{n}'}^{k}\right|=\Biggl|\sum_{n=1}^{N}Q_{n,a_{n}}^{k}-\sum_{n=1}^{N}Q_{n,a_{n}}+\sum_{n=1}^{N}Q_{n,a_{n}}-\sum_{n=1}^{N}Q_{n,a_{n}'}+\sum_{n=1}^{N}Q_{n,a_{n}'}-\sum_{n=1}^{N}Q_{n,a_{n}'}^{k}\Biggr|\underset{\left(a\right)}{\geq}\\
\left|\sum_{n=1}^{N}Q_{n,a_{n}}-\sum_{n=1}^{N}Q_{n,a_{n}'}\right|-\left|\sum_{n=1}^{N}Q_{n,a_{n}}^{k}-\sum_{n=1}^{N}Q_{n,a_{n}}\right|-\left|\sum_{n=1}^{N}Q_{n,a_{n}'}^{k}-\sum_{n=1}^{N}Q_{n,a_{n}'}\right|\underset{\left(b\right)}{>}\Delta_{\min}-\frac{3\Delta_{\min}}{4}\geq\frac{\Delta_{\min}}{4}\label{eq:24-1}
\end{multline}
where (a) follows from the reverse triangle inequality and (b) from
\eqref{eq:23-1} and $\underset{n,i}{\max}\left|Q_{n,i}^{k}-Q_{n,i}\right|<\frac{3\Delta_{\min}}{8N}$.

Denote by $\tilde{a}$ the allocation that the auction phase converges
to. If $\varepsilon<\frac{\Delta_{\min}}{4K}$, then Theorem 1 in
\cite{naparstek2013optimal} guarantees that
\begin{equation}
\left|\sum_{n=1}^{N}Q_{n,\tilde{a}_{n}}^{k}-\underset{a_{1},...,a_{N}}{\max}\sum_{n=1}^{N}Q_{n,a_{n}}^{k}\right|<\frac{\Delta_{\min}}{4K}N\leq\frac{\Delta_{\min}}{4}\label{eq:14}
\end{equation}
which, due to \eqref{eq:24-1}, is only possible if
\begin{equation}
\sum_{n=1}^{N}Q_{n,\tilde{a}_{n}}^{k}=\underset{a_{1},...,a_{N}}{\max}\sum_{n=1}^{N}Q_{n,a_{n}}^{k}\underset{\left(a\right)}{=}\arg\underset{a_{1},...,a_{N}}{\max}\sum_{n=1}^{N}Q_{n,a\left(n\right)}\label{eq:15}
\end{equation}
where (a) follows from Lemma \ref{lem:Percision} since we assume
that the $k$-th exploration phase succeeded.
\end{IEEEproof}
\bibliographystyle{IEEEtran}
\bibliography{csmabanditsarxiv}

\begin{thebibliography}{10}
\providecommand{\url}[1]{#1}
\csname url@samestyle\endcsname
\providecommand{\newblock}{\relax}
\providecommand{\bibinfo}[2]{#2}
\providecommand{\BIBentrySTDinterwordspacing}{\spaceskip=0pt\relax}
\providecommand{\BIBentryALTinterwordstretchfactor}{4}
\providecommand{\BIBentryALTinterwordspacing}{\spaceskip=\fontdimen2\font plus
\BIBentryALTinterwordstretchfactor\fontdimen3\font minus
  \fontdimen4\font\relax}
\providecommand{\BIBforeignlanguage}[2]{{%
\expandafter\ifx\csname l@#1\endcsname\relax
\typeout{** WARNING: IEEEtran.bst: No hyphenation pattern has been}%
\typeout{** loaded for the language `#1'. Using the pattern for}%
\typeout{** the default language instead.}%
\else
\language=\csname l@#1\endcsname
\fi
#2}}
\providecommand{\BIBdecl}{\relax}
\BIBdecl

\bibitem{Katzela2000}
I.~Katzela and M.~Naghshineh, ``Channel assignment schemes for cellular mobile
  telecommunication systems: A comprehensive survey,'' \emph{IEEE
  Communications Surveys Tutorials}, vol.~3, no.~2, pp. 10--31, Second 2000.

\bibitem{Chieochan2010}
S.~Chieochan, E.~Hossain, and J.~Diamond, ``Channel assignment schemes for
  infrastructure-based 802.11 {WLANs}: A survey,'' \emph{IEEE Communications
  Surveys Tutorials}, vol.~12, no.~1, pp. 124--136, First 2010.

\bibitem{Ku2015}
G.~Ku and J.~M. Walsh, ``Resource allocation and link adaptation in {LTE} and
  {LTE} advanced: A tutorial,'' \emph{IEEE Communications Surveys Tutorials},
  vol.~17, no.~3, pp. 1605--1633, thirdquarter 2015.

\bibitem{Tragos2013}
E.~Z. Tragos, S.~Zeadally, A.~G. Fragkiadakis, and V.~A. Siris, ``Spectrum
  assignment in cognitive radio networks: A comprehensive survey,'' \emph{IEEE
  Communications Surveys Tutorials}, vol.~15, no.~3, pp. 1108--1135, Third
  2013.

\bibitem{Tanab2017}
M.~E. Tanab and W.~Hamouda, ``Resource allocation for underlay cognitive radio
  networks: A survey,'' \emph{IEEE Communications Surveys Tutorials}, vol.~19,
  no.~2, pp. 1249--1276, Secondquarter 2017.

\bibitem{Wong1999}
C.~Y. Wong, R.~S. Cheng, K.~B. Lataief, and R.~D. Murch, ``Multiuser {OFDM}
  with adaptive subcarrier, bit, and power allocation,'' \emph{IEEE Journal on
  Selected Areas in Communications}, vol.~17, no.~10, pp. 1747--1758, Oct 1999.

\bibitem{Shen2005}
Z.~Shen, J.~G. Andrews, and B.~L. Evans, ``Adaptive resource allocation in
  multiuser {OFDM} systems with proportional rate constraints,'' \emph{IEEE
  Transactions on Wireless Communications}, vol.~4, no.~6, pp. 2726--2737, Nov
  2005.

\bibitem{Sadr2009}
S.~Sadr, A.~Anpalagan, and K.~Raahemifar, ``Radio resource allocation
  algorithms for the downlink of multiuser {OFDM} communication systems,''
  \emph{IEEE Communications Surveys Tutorials}, vol.~11, no.~3, pp. 92--106, rd
  2009.

\bibitem{Huberman2012}
S.~Huberman, C.~Leung, and T.~Le-Ngoc, ``Dynamic spectrum management ({DSM})
  algorithms for multi-user {xDSL},'' \emph{IEEE Communications Surveys
  Tutorials}, vol.~14, no.~1, pp. 109--130, First 2012.

\bibitem{Gao2008}
L.~Gao and S.~Cui, ``Efficient subcarrier, power, and rate allocation with
  fairness consideration for {OFDMA} uplink,'' \emph{IEEE Transactions on
  Wireless Communications}, vol.~7, no.~5, pp. 1507--1511, May 2008.

\bibitem{Huang2009}
J.~Huang, V.~G. Subramanian, R.~Agrawal, and R.~Berry, ``Joint scheduling and
  resource allocation in uplink {OFDM} systems for broadband wireless access
  networks,'' \emph{IEEE Journal on Selected Areas in Communications}, vol.~27,
  no.~2, pp. 226--234, February 2009.

\bibitem{Yaacoub2012}
E.~Yaacoub and Z.~Dawy, ``A survey on uplink resource allocation in {OFDMA}
  wireless networks,'' \emph{IEEE Communications Surveys Tutorials}, vol.~14,
  no.~2, pp. 322--337, Second 2012.

\bibitem{Papadimitriou1998}
{ C. Papadimitriou and K. Steiglitz}, \emph{Combinatorial Optimization:
  Algorithms and Complexity}.\hskip 1em plus 0.5em minus 0.4em\relax Dover,
  1998.

\bibitem{bertsekas1979distributed}
\BIBentryALTinterwordspacing
D.~P. Bertsekas, ``The auction algorithm: A distributed relaxation method for
  the assignment problem,'' \emph{Annals of Operations Research}, vol.~14,
  no.~1, pp. 105--123, Dec 1988. [Online]. Available:
  \url{https://doi.org/10.1007/BF02186476}
\BIBentrySTDinterwordspacing

\bibitem{naparstek2013optimal}
O.~Naparstek and A.~Leshem, ``Fully distributed optimal channel assignment for
  open spectrum access,'' \emph{IEEE Transactions on Signal Processing},
  vol.~62, no.~2, pp. 283--294, Jan 2014.

\bibitem{Challita2017}
U.~Challita, L.~Dong, and W.~Saad, ``Proactive resource management in {LTE-U}
  systems: A deep learning perspective,'' \emph{arXiv preprint
  arXiv:1702.07031}, June 2017.

\bibitem{Wang2018}
S.~Wang, H.~Liu, P.~H. Gomes, and B.~Krishnamachari, ``Deep reinforcement
  learning for dynamic multichannel access in wireless networks,'' \emph{IEEE
  Transactions on Cognitive Communications and Networking}, vol.~4, no.~2, pp.
  257--265, June 2018.

\bibitem{Naparstek}
O.~Naparstek and K.~Cohen, ``Deep multi-user reinforcement learning for
  distributed dynamic spectrum access,'' arXiv preprint arXiv:1704.02613, 2017.

\bibitem{Nayyar2016}
N.~Nayyar, D.~Kalathil, and R.~Jain, ``On regret-optimal learning in
  decentralized multi-player multi-armed bandits,'' \emph{IEEE Transactions on
  Control of Network Systems}, vol.~5, no.~1, pp. 597--606, 2018.

\bibitem{Avner2014}
O.~Avner and S.~Mannor, ``Concurrent bandits and cognitive radio networks,'' in
  \emph{Joint European Conference on Machine Learning and Knowledge Discovery
  in Databases}, 2014, pp. 66--81.

\bibitem{Anandkumar2011}
A.~Anandkumar, N.~Michael, A.~K. Tang, and A.~Swami, ``Distributed algorithms
  for learning and cognitive medium access with logarithmic regret,''
  \emph{IEEE Journal on Selected Areas in Communications}, vol.~29, no.~4, pp.
  731--745, 2011.

\bibitem{Liu2012}
K.~Liu and Q.~Zhao, ``Cooperative game in dynamic spectrum access with unknown
  model and imperfect sensing,'' \emph{IEEE Transactions on Wireless
  Communications}, vol.~11, no.~4, pp. 1596--1604, April 2012.

\bibitem{Liu2010Restless}
------, ``Indexability of restless bandit problems and optimality of whittle
  index for dynamic multichannel access,'' \emph{IEEE Transactions on
  Information Theory}, vol.~56, no.~11, pp. 5547--5567, Nov 2010.

\bibitem{Vakili2013}
S.~Vakili, K.~Liu, and Q.~Zhao, ``Deterministic sequencing of exploration and
  exploitation for multi-armed bandit problems,'' \emph{IEEE Journal of
  Selected Topics in Signal Processing}, vol.~7, no.~5, pp. 759--767, 2013.

\bibitem{Rosenski2016}
J.~Rosenski, O.~Shamir, and L.~Szlak, ``Multi-player bandits--a musical chairs
  approach,'' in \emph{International Conference on Machine Learning}, 2016, pp.
  155--163.

\bibitem{Liu2010MAB}
K.~Liu and Q.~Zhao, ``Distributed learning in multi-armed bandit with multiple
  players,'' \emph{IEEE Transactions on Signal Processing}, vol.~58, no.~11,
  pp. 5667--5681, 2010.

\bibitem{Kalathil2014}
D.~Kalathil, N.~Nayyar, and R.~Jain, ``Decentralized learning for multiplayer
  multiarmed bandits,'' \emph{IEEE Transactions on Information Theory},
  vol.~60, no.~4, pp. 2331--2345, 2014.

\bibitem{Bistritz}
I.~Bistritz and A.~Leshem, ``Game of thrones: Fully distributed learning for
  multi-player bandits,'' arXiv:1810.11162.

\bibitem{bistritz2018distributed}
------, ``Distributed multi-player bandits-a game of thrones approach,'' in
  \emph{Advances in Neural Information Processing Systems}, 2018, pp.
  7222--7232.

\bibitem{Kwon2010}
H.~Kwon, S.~Kim, and B.~G. Lee, ``Opportunistic multi-channel {CSMA} protocol
  for {OFDMA} systems,'' \emph{IEEE Transactions on Wireless Communications},
  vol.~9, no.~5, pp. 1552--1557, May 2010.

\bibitem{Yaffe2010}
Y.~Yaffe, A.~Leshem, and E.~Zehavi, ``Stable matching for channel access
  control in cognitive radio systems,'' in \emph{2010 2nd International
  Workshop on Cognitive Information Processing}, June 2010, pp. 470--475.

\bibitem{Leshem2012}
A.~Leshem, E.~Zehavi, and Y.~Yaffe, ``Multichannel opportunistic carrier
  sensing for stable channel access control in cognitive radio systems,''
  \emph{IEEE Journal on Selected Areas in Communications}, vol.~30, no.~1, pp.
  82--95, January 2012.

\bibitem{gale1962college}
D.~Gale and L.Shapley, ``College admissions and the stability of marriage,''
  \emph{The American Mathematical Monthly}, vol.~69, no.~1, pp. 9--15, 1962.

\bibitem{Naparstek2012}
O.~Naparstek and A.~Leshem, ``Bounds on the expected optimal channel assignment
  in {R}ayleigh channels,'' in \emph{2012 IEEE 13th International Workshop on
  Signal Processing Advances in Wireless Communications (SPAWC)}, June 2012,
  pp. 294--298.

\bibitem{Mochaourab2015}
R.~Mochaourab, B.~Holfeld, and T.~Wirth, ``Distributed channel assignment in
  cognitive radio networks: Stable matching and {W}alrasian equilibrium,''
  \emph{IEEE Transactions on Wireless Communications}, vol.~14, no.~7, pp.
  3924--3936, July 2015.

\bibitem{Bistritz2018}
I.~Bistritz and A.~Leshem, ``Game theoretic dynamic channel allocation for
  frequency-selective interference channels,'' \emph{IEEE Transactions on
  Information Theory}, 2018, {DOI}: 10.1109/TIT.2018.2868440.

\bibitem{Xiao2015}
Y.~Xiao, K.~C. Chen, C.~Yuen, Z.~Han, and L.~A. DaSilva, ``A bayesian
  overlapping coalition formation game for device-to-device spectrum sharing in
  cellular networks,'' \emph{IEEE Transactions on Wireless Communications},
  vol.~14, no.~7, pp. 4034--4051, July 2015.

\bibitem{leshem20112}
E.~Zehavi and A.~Leshem, ``Bargaining solution for partial orthogonal
  transmission over frequency selective interference channel,'' in \emph{2011
  IEEE International Symposium on Information Theory Proceedings}, July 2011,
  pp. 2701--2705.

\bibitem{Han2005}
Z.~Han, Z.~Ji, and K.~J.~R. Liu, ``Fair multiuser channel allocation for
  {OFDMA} networks using {N}ash bargaining solutions and coalitions,''
  \emph{IEEE Transactions on Communications}, vol.~53, no.~8, pp. 1366--1376,
  Aug 2005.

\bibitem{Leshem2011}
A.~Leshem and E.~Zehavi, ``Smart carrier sensing for distributed computation of
  the generalized nash bargaining solution,'' in \emph{2011 17th International
  Conference on Digital Signal Processing (DSP)}, July 2011, pp. 1--5.

\bibitem{Cohen2013Journal}
K.~Cohen, A.~Leshem, and E.~Zehavi, ``Game theoretic aspects of the
  multi-channel {ALOHA} protocol in cognitive radio networks,'' \emph{IEEE
  Journal on Selected Areas in Communications}, vol.~31, no.~11, pp.
  2276--2288, November 2013.

\bibitem{Cohen2013}
K.~Cohen and A.~Leshem, ``Distributed throughput maximization for multi-channel
  {ALOHA} networks,'' in \emph{2013 5th IEEE International Workshop on
  Computational Advances in Multi-Sensor Adaptive Processing (CAMSAP)}, Dec
  2013, pp. 456--459.

\bibitem{Sun2006}
J.~Sun, E.~Modiano, and L.~Zheng, ``Wireless channel allocation using an
  auction algorithm,'' \emph{IEEE Journal on Selected Areas in Communications},
  vol.~24, no.~5, pp. 1085--1096, May 2006.

\bibitem{Han2011}
Z.~Han, R.~Zheng, and H.~V. Poor, ``Repeated auctions with {B}ayesian
  nonparametric learning for spectrum access in cognitive radio networks,''
  \emph{IEEE Transactions on Wireless Communications}, vol.~10, no.~3, pp.
  890--900, March 2011.

\bibitem{Mukherjee2010}
A.~Mukherjee and H.~M. Kwon, ``General auction-theoretic strategies for
  distributed partner selection in cooperative wireless networks,'' \emph{IEEE
  Transactions on Communications}, vol.~58, no.~10, pp. 2903--2915, October
  2010.

\bibitem{Chang2010}
H.~B. Chang and K.~C. Chen, ``Auction-based spectrum management of cognitive
  radio networks,'' \emph{IEEE Transactions on Vehicular Technology}, vol.~59,
  no.~4, pp. 1923--1935, May 2010.

\bibitem{Yang2009}
K.~Yang, N.~Prasad, and X.~Wang, ``An auction approach to resource allocation
  in uplink {OFDMA} systems,'' \emph{IEEE Transactions on Signal Processing},
  vol.~57, no.~11, pp. 4482--4496, Nov 2009.

\bibitem{Bayati2007}
M.~Bayati, B.~Prabhakar, D.~Shah, and M.~Sharma, ``Iterative scheduling
  algorithms,'' in \emph{IEEE INFOCOM 2007 - 26th IEEE International Conference
  on Computer Communications}, May 2007, pp. 445--453.

\bibitem{Bayati2008}
M.~Bayati, D.~Shah, and M.~Sharma, ``Max-product for maximum weight matching:
  Convergence, correctness, and {LP} duality,'' \emph{IEEE Transactions on
  Information Theory}, vol.~54, no.~3, pp. 1241--1251, March 2008.

\bibitem{Naparstek2013}
O.~Naparstek and A.~Leshem, ``A fast matching algorithm for asymptotically
  optimal distributed channel assignment,'' in \emph{2013 18th International
  Conference on Digital Signal Processing (DSP)}, July 2013, pp. 1--6.

\bibitem{Meredith2016}
J.~Meredith, ``Study on channel model for frequency spectrum above 6 ghz,''
  3GPP TR 38.900, Jun, Tech. Rep., 2016.

\bibitem{Sun2016}
S.~Sun, T.~S. Rappaport, S.~Rangan, T.~A. Thomas, A.~Ghosh, I.~Z. Kovacs,
  I.~Rodriguez, O.~Koymen, A.~Partyka, and J.~Jarvelainen, ``Propagation path
  loss models for 5g urban micro-and macro-cellular scenarios,'' in
  \emph{Vehicular Technology Conference (VTC Spring), 2016 IEEE 83rd}.\hskip
  1em plus 0.5em minus 0.4em\relax IEEE, 2016, pp. 1--6.

\bibitem{Hoeffding1963}
W.~Hoeffding, ``Probability inequalities for sums of bounded random
  variables,'' \emph{Journal of the American Statistical Association}, vol.~58,
  no. 301, pp. 13--30, 1963.

\end{thebibliography}

\end{document}